\documentclass{WileyMSP-template}
\usepackage{amsmath,amsfonts}
\usepackage[numbers,sort&compress]{natbib}
\usepackage{xcolor}
\definecolor{mygray}{gray}{0.9}
\usepackage{parskip}
\PassOptionsToPackage{a4paper,margin=20cm}{geometry}
\usepackage{geometry}
\usepackage{ragged2e}
\usepackage{hyperref}
\usepackage{lineno} 
\usepackage{relsize}
\justifying

\begin{document}
\pagestyle{fancy}

\title{Inverse-designed dispersive time-varying nanostructures}
\maketitle


\author{Puneet~Garg*,}
\author{Jan~David~Fischbach,}
\author{Aristeidis G. Lamprianidis,}
\author{Xuchen~Wang,}\\
\author{Mohammad~S.~Mirmoosa,}
\author{Viktar~S.~Asadchy,}
\author{Carsten~Rockstuhl*,}
\author{ and Thomas~J.~Sturges*}

\begin{affiliations}
P.~Garg, Dr.~A.~G.~Lamprianidis, Prof.~C.~Rockstuhl\\
Institute of Theoretical Solid State Physics, Karlsruhe Institute of Technology, Kaiserstr. 12, 76131 Karlsruhe, Germany\\
Email Address: puneet.garg@kit.edu; carsten.rockstuhl@kit.edu\\
J.~D.~Fischbach, Prof.~C.~Rockstuhl, Dr.~T.~J.~Sturges\\
Institute of Nanotechnology, Karlsruhe Institute of Technology, Kaiserstr. 12, 76131 Karlsruhe, Germany\\
Email Address: thomas.sturges@kit.edu\\
Dr.~X.~Wang\\
Qingdao Innovation and Development Base, Harbin Engineering University, Qingdao 266400, China\\
Dr.~M.~S.~Mirmoosa\\
Department of Physics and Mathematics, University of Eastern Finland, Joensuu, Finland\\
Prof.~V.~S.~Asadchy\\
Department of Electronics and Nanoengineering, Aalto University, Espoo, Finland

\end{affiliations}



\begin{abstract}
Time-varying nanostructures allow us to control the spatial and temporal properties of light. The temporal modulation of the nanostructures constitutes an additional degree of freedom to control their scattering properties on demand and in a reconfigurable manner. However, these additional parameters create a vast design space, raising challenges in identifying optimal designs. Therefore, tools from the field of photonic inverse design must be used to optimize the degrees of freedom of the system to facilitate predefined optical responses. To further develop this field, here we introduce a differentiable transition (T-) matrix-based inverse design framework for dispersive time-varying nanostructures. The electron density of the material of the nanostructures is modulated non-adiabatically as a generic periodic function of time. Using the inverse design framework, the temporal shape of the electron density can be manipulated to reach the target functionality. Our computational framework is exploited, exemplarily, in two instances. First, the decay rate enhancement of oscillating dipoles near time-varying spheres is controlled on demand. Second, using spatiotemporal metasurfaces, a system supporting asymmetric transmission of light at visible frequencies is designed. Our work paves the way toward programmable spatiotemporal metasurfaces and space-time crystals for a future generation of reconfigurable functional photonic devices.  
\end{abstract}


\section{Introduction}

Time-varying nanostructures facilitate the simultaneous manipulation of light's spatial and temporal properties \cite{engheta2023four}. In this context, time modulation refers to changing the material properties as a function of time~\cite{galiffi2022photonics,ptitcyn2023tutorial,asgari2024photonic}. Additionally, in the case of nanostructures, we consider spatially localized structured objects. Recently, various applications exploiting time-varying media have been reported. Some of them are made from spatially homogeneous, and some of them are made from spatially structured materials. These applications include parametric amplification \cite{zurita2009reflection,wang2023unleashing}, nonreciprocal light propagation \cite{shaltout2015time,huidobro2021homogenization}, asymmetric frequency conversion \cite{mirmoosa2024time}, power combining of waves \cite{wang2021space}, temporal aiming~\cite{pacheco2020temporal}, inverse prisms \cite{akbarzadeh2018inverse}, wave freezing~\cite{wang2023controlling}, anti-reflection temporal coatings~\cite{ramaccia2020light,pacheco2020antireflection}, perfect absorption \cite{mostafa2022coherently,hayran2024beyond}, polarization conversion \cite{yang2021simultaneous,mostafa2023spin,mirmoosa2024time}, spatiotemporal wavefront shaping \cite{globosits2024photonic}, and so forth. Besides, various experimental advancements in the field of time-varying materials have fueled the interest of the photonics community in this research domain \cite{Lustig2023time,tirole2023double,Moussa2023observation,wang2023metasurface,galiffi2023broadband,sisler2024electrically,harwood2024super}. Owing to the tunability of the temporal modulation, several reconfigurable devices with desired functionalities can be realized using such dynamic materials \cite{buddhiraju2021arbitrary,sisler2024electrically,zhang2024coprime}. It has to be emphasized that the ability to drastically change the optical response of photonic structures after fabrication is in striking contrast to most conventional systems whose properties are fixed upon fabrication. Of course, multiple other approaches exist to tune the optical response after fabrication, but a temporal modulation seems to be particularly versatile.   

Generally, the temporal modulation unlocks additional degrees of freedom to control the flow of light. However, these additional parameters are a blessing and a curse in that they create a vast design space but simultaneously challenge the identification of optimal designs. It prompts the use of tools from the field of photonic inverse design, a framework well established by now to design optical devices with tailored functionality \cite{Molesky2018inverse,sturges2024inverse,kuster2024inverse,zhu2023inverse}. Generally, photonic inverse design is used to iteratively optimize the parameters of a nanostructure toward a design that fulfills a predefined optical functionality \cite{Molesky2018inverse}. It frequently utilizes fully differentiable software tools that incorporate built-in adjoint solvers, allowing for the optimization of devices with numerous design parameters through a gradient-based method.

Photonic inverse design has already been used to engineer the interaction of light with time-varying systems. Examples include optical pulse shaping for their enhanced interaction with temporally periodic media \cite{Baxter2023inverse}, unidirectional scattering from disordered photonic time crystals \cite{kim2023unidirectional}, topological state design for photonic time crystals \cite{Long2024inverse}, non-reciprocal light propagation from spatiotemporal nanostructures \cite{phi2024controlled}, and spectral control of light from time-varying dispersionless spheres \cite{sadafi2023dynamic}. It is important to remark that most materials, whose properties can be modulated as a function of time with modulation frequencies comparable to the oscillation frequency of light (i.e., non-adiabatic modulation), exhibit dispersion in the optical regime \cite{lobet2023new,horsley2023eigenpulses,baxter2023dynamic}. However, to the best of the authors' knowledge, a study involving the photonic inverse design of dispersive and non-adiabatically modulated optical nanostructures has not yet been reported. 

To effectively use the techniques from the field of photonic inverse design, the forward problem needs to be solved efficiently at first. 
Using traditional full-wave Maxwell solvers to model such nanostructures, e.g., based on the finite-element method, is often computationally expensive \cite{ptitcyn2023floquet,garg2022modeling}. However, a particularly useful formalism to model such nanostructures efficiently is the transition matrix (also known as the T-matrix) formalism \cite{waterman1965proceedings}. The T-matrix encapsulates the linear scattering response of the considered nanostructure. Once the T-matrix is known, we can calculate the field scattered off the nanostructure for an arbitrary illumination. In fact, given the T-matrix of a finite scatterer, one can compute the effective T-matrix of an infinite periodic arrangement of such a scatterer using the Ewald summation technique \cite{ewald1921die,beutel2024treams}. That approach allows us to study the optical response from spatiotemporal metasurfaces. Furthermore, the T-matrix model can be made differentiable, enabling gradient-based optimization techniques to be applied to designing the underlying nanostructures.

In this article, we propose a differentiable T-matrix-based inverse design framework that incorporates material dispersion and non-adiabatic temporal modulations. Our framework can handle time-varying spheres and spatiotemporal metasurfaces made from such spheres. We assume the underlying material of the spheres to have a generic time-periodic electron density. We model these time-varying nanostructures using the differentiable T-matrix formalism. Moreover, we use a gradient-based optimization technique to tailor the scattering characteristics of the nanostructures on demand. In particular, we optimize the temporal profile of the electron density of the underlying material to reach the desired optical responses for the nanostructures. 
Utilizing the proposed inverse design framework, we showcase its applicability in two key examples. First, we design the electron density of the material of a sphere such that a dipole near it exhibits anomalous Drexhage effect. Second, we design the electron density of the material of a spatiotemporal metasurface that leads to an efficient frequency upconversion in transmission. This metasurface is then cascaded with a narrow-band frequency filter to realize the asymmetric transmission of light. Importantly, the parameters of the system are chosen so that asymmetric transmission is achieved for visible light.

\section{Inverse design framework}\label{sec:inverse}
\begin{figure*}
\centerline{\includegraphics[width= 0.7\columnwidth]{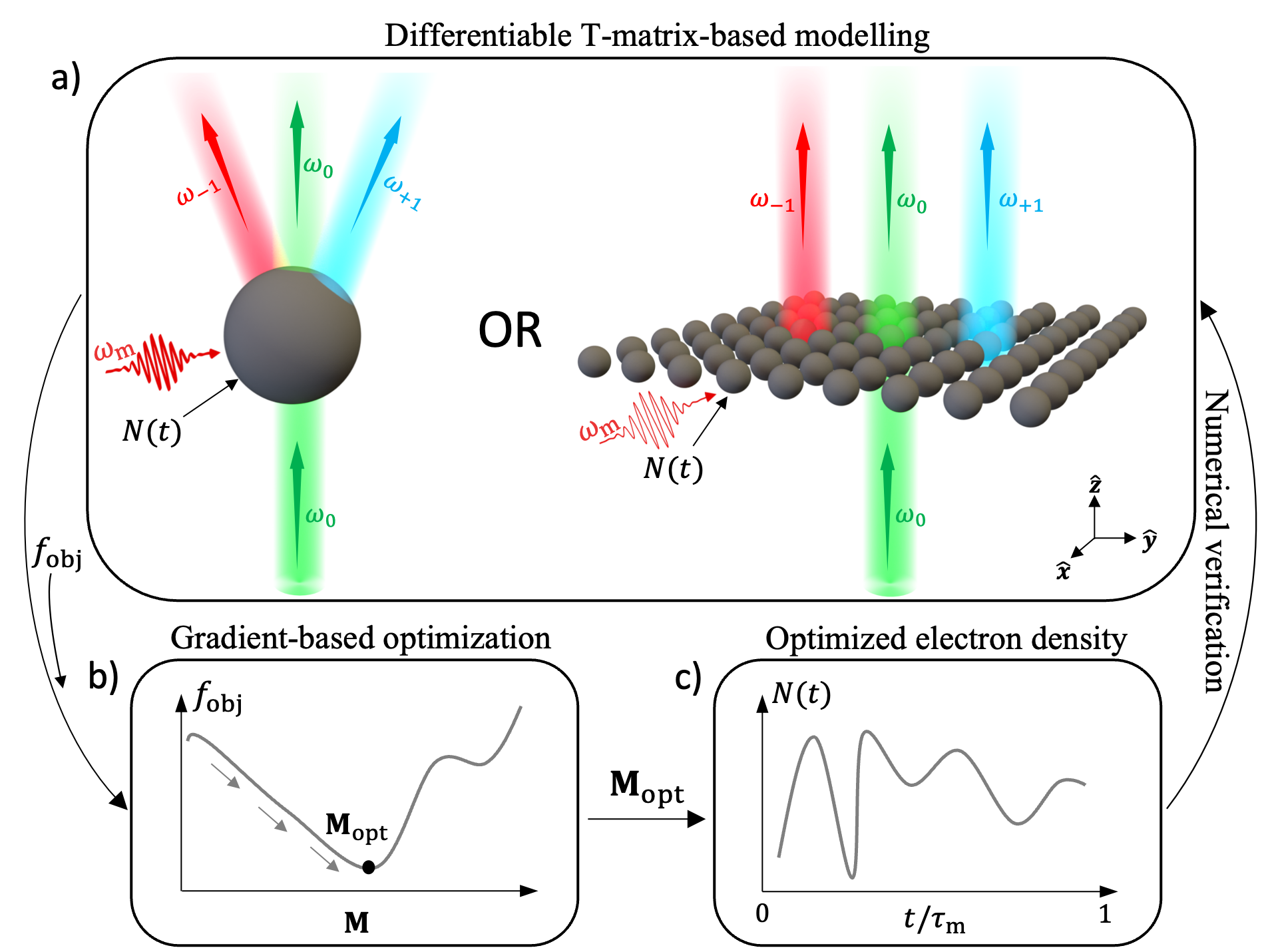}}
\caption{Inverse design framework. First, a time-varying sphere or spatiotemporal metasurface, as shown in (a), is numerically set up using the differentiable T-matrix-based model. Then, depending on the desired functionality of the considered nanostructure, an objective function $f_\mathrm{obj}(\mathbf{M})$ is defined. The time variance of the nanostructure is encoded in the Fourier coefficients $\mathbf{M}$ of the electron density $N(t)$. Next, $f_\mathrm{obj}$ is minimized or maximized using gradient-based optimization as shown in (b). The optimization returns the optimal Fourier coefficients $\mathbf{M}\,\,{=}\,\,\mathbf{M}_\mathrm{opt}$ that correspond to an optimal time-varying electron density $N(t)$ shown in (c). Finally, the optimization results are numerically verified for the desired functionality by a forward simulation of the considered nanostructure having the optimal $N(t)$. Here, $\tau_\mathrm{m}\,\,{=}\,\,2\pi/\omega_\mathrm{m}$ corresponds to the period of the temporal modulation.}
\label{fig:concept}
\end{figure*}
In this section, the optimization framework used to design the time-varying nanostructures is presented (see \textbf{Figure~\ref{fig:concept}}). In principle, there are many possible approaches. Global optimization techniques such as Bayesian optimization \cite{Frazier2018bayesian} and particle swarm optimization \cite{kennedy1995particle} constitute powerful tools for optimizing complicated objective functions (such as ours). However, they scale quite poorly with the number of design parameters. Deep learning approaches are also possible \cite{yu2020machine,ma2021deep}. However, the computational costs of generating a suitable training dataset are relatively high, making this approach undesirable in situations where only a few individual optimizations are needed. For all these reasons, we choose automatic differentiation in combination with a local gradient-based optimization algorithm to design our structures. This autodiff approach allows us to efficiently find optimized designs with a computational cost that is independent of the number of design variables, making this a very scalable approach. While there is no guarantee of converging to the global optimum, many locally optimal solutions often yield satisfactory designs. Consequently, for practical purposes, local gradient-based optimization serves as an efficient and effective tool for inverse design. We note that our approach is conceptually similar to ``adjoint optimization'' \cite{yeung2022enhancing,hughes2018adjoint}. However, our autodiff approach gives us the exact gradients of the discretized system, at the cost of a significantly higher memory requirement. We begin by outlining the differentiable T-matrix method used for modeling time-varying nanostructures. Then, we describe the gradient-based optimization procedure.

\subsection{Differentiable T-matrix method}
In this article, we optimize the optical properties of time-varying spheres and spatiotemporal metasurfaces (see Figure~\ref{fig:concept}(a)). The considered spatiotemporal metasurfaces are periodic arrays of time-varying spheres arranged in two dimensions (2D). We assume these nanostructures to be made from a dispersive material whose electron density $N(t)$ is a periodic function of time, with a period of $\tau_\mathrm{m}\,\,{=}\,\,2\pi/\omega_\mathrm{m}$. Here, we consider the non-adiabatic modulation regime, where $\omega_\mathrm{m}$ is comparable to the oscillation frequency of light. Such a time-varying electron density $N(t)$ is incorporated into our analysis using the Drude dispersion model \cite{ptitcyn2023floquet, garg2022modeling} (see Supporting Information; Section~S1). Note that for all the examples in this manuscript, we use Drude parameters that resemble the properties of gold at frequencies below its plasma frequency \cite{blaber2009search}. Such a choice is made as gold exhibits strong dispersion at visible frequencies that are of interest in this work. Importantly, due to the time modulation of the nanostructures, an incident excitation with frequency $\omega$ gives rise to a polychromatic scattered field with frequencies $\omega_j=\omega+j\omega_\mathrm{m}$, with $j\in \mathbb{Z}$ (see Figure~\ref{fig:concept}(a)). 

As mentioned earlier, linear optical nanostructures can be numerically modeled using the T-matrix method \cite{waterman1965proceedings}. The T-matrix is an efficient tool to semi-analytically model optical nanostructures based on Mie theory \cite{Mie1908beitrage}. It can handle finite scatterers and photonic materials made from an infinite periodic arrangement of such scatterers. The T-matrix formalism involves expanding the incident field $\mathbf{E}^\mathrm{inc}$ and scattered field $\mathbf{E}^\mathrm{sca}$ for the nanostructure in a basis of vector spherical harmonics (see the Methods section). Let $\mathbf{A}^\mathrm{inc}$ and $\mathbf{A}^\mathrm{sca}$ be the corresponding incident and scattered coefficient vectors, respectively. Then, $\mathbf{A}^\mathrm{inc}$ and $\mathbf{A}^\mathrm{sca}$ satisfy
\begin{eqnarray}
\mathbf{A}^\mathrm{sca}= \mathbf{\hat{T}}\cdot\mathbf{A}^\mathrm{inc}\, ,\label{eq:Tmat}
\end{eqnarray}
where $\mathbf{\hat{T}}$ is the so-called T-matrix of the considered nanostructure. From Equation~\eqref{eq:Tmat}, we note that the T-matrix completely encapsulates the scattering response of the nanostructure. Once the T-matrix of the nanostructure is known, the field scattered off it can be calculated easily using Equation~\eqref{eq:Tmat}. The T-matrix for static scatterers can be computed analytically or using full-wave Maxwell solvers \cite{comsol,JCM,asadova2024tmatrix}. Note that the T-matrix formalism is closely related to the widely used scattering matrix (S-matrix) formalism. In fact, once the T-matrix of a nanostructure is known, its S-matrix can be calculated by performing appropriate basis transformations as shown in the Supporting Information in Section~S2. Recently, the T-matrix formalism has also been applied to time-varying nanostructures \cite{ptitcyn2023floquet,garg2022modeling,Stefanou2023light,stefanou2021light,panagiotidis2023optical,sadafi2023dynamic}. The T-matrices of dispersive time-varying spheres and spatiotemporal metasurfaces made from these time-varying spheres are analytically known \cite{ptitcyn2023floquet,garg2022modeling}. We use these existing T-matrices for our purposes. We wish to highlight here that Equation~\eqref{eq:Tmat} also holds for the time-varying nanostructures. In that case, the size of $\mathbf{\hat{T}}$ depends on the number of scattered frequencies $\omega_j$ and considered vector spherical harmonics (see the Methods section). Furthermore, it is important to mention that we wrote our T-matrix-based code using JAX, a software package that can automatically differentiate native Python and Numpy functions \cite{jax2018github}. Therefore, it makes our T-matrix model differentiable, allowing us to use a gradient-based optimization as discussed in the following.

\subsection{Optimization procedure}
Having established the differentiable T-matrix-based model, we next discuss the optimization procedure. We use a gradient-based optimization approach. We begin by expanding $N(t)$ in a Fourier series. Such an expansion of $N(t)$ in a Fourier series is done as we solve the problem in the frequency domain in our computational framework. Therefore, $N(t)$ can be written as
\begin{equation}\label{eq:N(t)}
    N(t)=N_0\left[1+\sum_{q=1}^{Q}(M_q e^{-iq\omega_\mathrm{m}t}+\textrm{c.c.})\right]\,.
\end{equation}
Here, $N_0$ corresponds to the electron density of the static material, the different $M_q$ correspond to the complex Fourier coefficients of $N(t)$, $Q$ is the total number of considered $M_q$ coefficients, and $\textrm{c.c.}$ refers to the complex conjugate. The electron density $N(t)$ is non-negative and real-valued. We define a vector $\mathbf{M}$ that contains all the coefficients $M_q$ for brevity. Then, depending on the desired functionality of the time-varying nanostructure, an objective function $f_\mathrm{obj}(\mathbf{M})$ is defined. The objective function $f_\mathrm{obj}$ takes $\mathbf{M}$ as the design variables. Note that the base modulation frequency $\omega_\mathrm{m}$ is chosen \textit{a priori} and does not enter the optimization procedure.

Next, depending on the design goal, $f_\mathrm{obj}$ is either minimized or maximized using gradient-based optimization (see Figure~1(b)). As mentioned earlier, the differentiable T-matrix-based model allows us to compute the derivatives of $f_\mathrm{obj}$ with respect to $\mathbf{M}$. Additionally, thanks to the reverse-mode automatic differentiation, we obtain the derivatives of $f_\mathrm{obj}$ with respect to all $M_q \in \mathbf{M}$ with a single backward pass through the simulation \cite{margossian2019areview}. More details on the software packages can be found in Supporting Information; Subsection~S3.1. Using the gradient information, we iteratively update $\mathbf{M}$ until a local minimum/maximum of $f_\mathrm{obj}$ is reached (see Figure~\ref{fig:concept}(b)). Such an optimization returns the optimal Fourier coefficient vector $\mathbf{M}=\mathbf{M}_\mathrm{opt}$, and correspondingly, the optimal time-varying electron density $N(t)$ (see Figure~\ref{fig:concept}(c)). Finally, we numerically verify if the desired functionality is reached by running a forward simulation with the optimal $N(t)$ (see Figure~\ref{fig:concept}(a)). Note that the details of various convergence parameters used for calculations in this manuscript are available in the Supporting Information.  

Having introduced the inverse design framework, we now exploit it to study two distinct phenomena. First, we tailor the electron density of the material of a time-varying sphere so that a dipole in its vicinity displays anomalous Drexhage effect. Next, we present an example based on the spatiotemporal metasurfaces to realize asymmetric transmission of light at visible frequencies.

\section{Radiative decay rate enhancement}
Nanostructures are extensively used in the photonics community to enhance the spontaneous radiative decay rates of excited atoms \cite{pelton2015modified,stamatopoulou2021role,elli2023strong}. In classical theory, an excited atom can be modeled as an oscillating point electric dipole \cite{giannini2011plasmonic}. Recently, there has been a growing interest in studying the spontaneous decay rates of dipoles in the vicinity of time-varying media \cite{lyuborav2022amplified,calajo2017control,park2024spontaneous,Yu2023manipulating}. Time-varying media are non-Hermitian systems, as they can supply (leading to optical gain) and absorb (leading to optical loss) energy to and from the incident electromagnetic fields. However, optical gain necessitates a careful examination of the existing approaches to calculate the decay rates of dipoles near such non-Hermitian systems \cite{franke2021fermi,ren2024classical}. Therefore, for the dipoles in the vicinity of time-varying nanostructures, the effects of optical gain on their spontaneous decay rate calculation must be considered. The optical gain can be interpreted as a spontaneous excitation of the atom from its ground to an excited state, eventually leading to the subsequent emission of a photon \cite{franke2021fermi,park2024spontaneous}. Here, the excitation of the atom is not photon-mediated; rather it occurs due to the non-adiabatic quantum pumping induced by the time-periodic material \cite{franke2021fermi,park2024spontaneous}. Note that non-adiabatic quantum pumping is a process where a system, due to rapid time modulation, cannot respond instantaneously and consequently ceases to remain in its ground state \cite{citro2023thouless}. The calculation of the decay rates of dipoles in the vicinity of time-periodic media has been shown in various works \cite{lyuborav2022amplified,calajo2017control,park2024spontaneous,Yu2023manipulating}. However, a computation of the decay rates of dipoles kept near spatially finite time-varying media has not yet been presented. Finite media are advantageous as they allow greater flexibility in positioning the dipoles near them. In the following, using the arguments of \cite{ren2024classical}, we derive an expression for the radiative decay rate enhancement of a dipole kept near a time-varying sphere while considering the optical gain (and loss) from the sphere.

\subsection{Derivation of the radiative decay rate enhancement}
\begin{figure*}
\centerline{\includegraphics[width= 0.75\columnwidth,trim=1 1 0.5 1,clip]{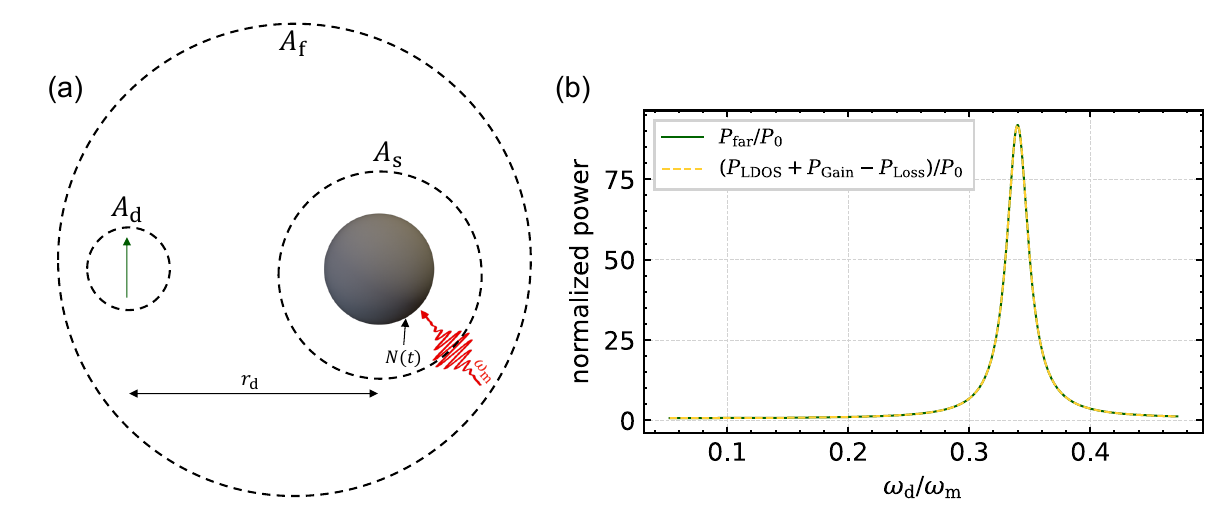}}
\caption{(a) A system consisting of an electric dipole near a time-varying sphere. (b) Normalized powers plotted as a function of the emission frequency $\omega_\mathrm{d}$ of the dipole for the system shown in (a). Here, the time-varying electron density of the sphere is given by $N(t)=N_0[1+0.8\mathrm{cos}(\omega_\mathrm{m}t)]$, and its radius is $R=10$~nm. Furthermore, the dipole is kept at a distance $r_\mathrm{d}=15$~nm from the center of the sphere. Moreover, $\omega_\mathrm{m}=0.17c_0/R$ with $c_0$ being the speed of light in vacuum.}
\label{fig:power_balance}
\end{figure*}
\begin{figure*}
\centerline{\includegraphics[width= 0.8\columnwidth,trim=1 1 0.5 1,clip]{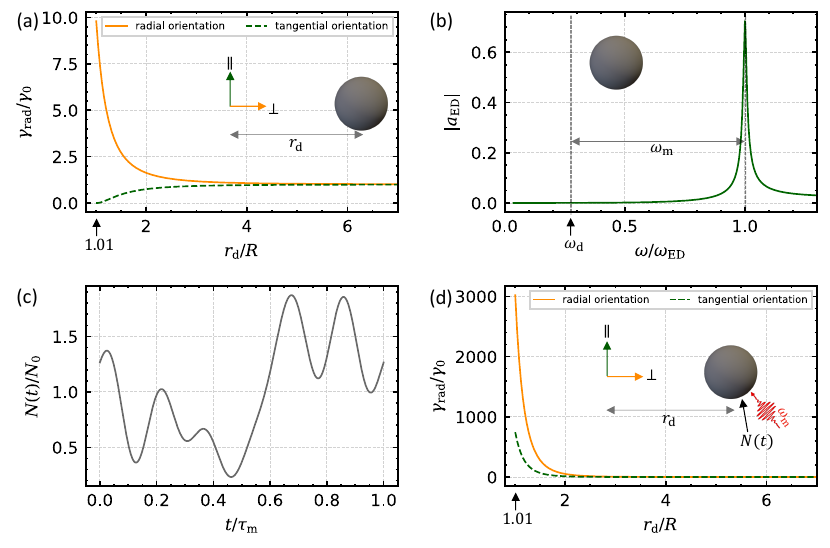}}
\caption{Anomalous Drexhage Effect. (a) The radiative decay rate enhancement ${\gamma_\mathrm{rad}}/{\gamma_\mathrm{0}}$ for an electric dipole as a function of its distance $r_\mathrm{d}$ from the center of a static sphere for its two orthogonal polarizations. (b) The absolute value of the electric dipolar Mie coefficient of the static sphere as a function of the excitation frequency $\omega$. (c) The optimized electron density $N(t)$ that maximizes ${\gamma_\mathrm{rad}}/{\gamma_\mathrm{0}}$ for the tangentially orientated dipole ($\parallel$- polarization)} at $r_\mathrm{d}/R=1.01$. (d) The radiative decay rate enhancement ${\gamma_\mathrm{rad}}/{\gamma_\mathrm{0}}$ of the dipole in two orthogonal polarizations as a function of its distance $r_\mathrm{d}$ from the time-varying sphere having the optimized $N(t)$ shown in (c).
\label{fig:drexhage}
\end{figure*}
The power conservation relation for a dipole kept in the vicinity of an arbitrary active (gain providing) and an arbitrary absorbing resonator reads as \cite[Equation~(21)]{ren2024classical}
\begin{eqnarray}
P_\mathrm{far}= P_\mathrm{LDOS}+P_\mathrm{gain}-P_\mathrm{abs}\,.
\label{eq:balance}
\end{eqnarray}
Here, $P_\mathrm{far}$ refers to the total power radiated by the system formed by the dipole and the resonators in the far field, $P_\mathrm{LDOS}$ refers to the net power crossing an imaginary surface enclosing only the dipole, $P_\mathrm{gain}$ refers to the total power gained by the system due to the active resonator, and $P_\mathrm{abs}$ refers to the total power absorbed by the lossy resonator from the system. Note that, in our analysis, time-averaged powers are considered (see the Supporting Information; Section~S5).

We aim to derive an expression for the radiative decay rate enhancement of a dipole kept near a time-varying sphere (see \textbf{Figure~\ref{fig:power_balance}}(a)). As mentioned earlier, a time-varying sphere acts as a medium that can supply and absorb energy to and from the fields incident on it. As Equation~\eqref{eq:balance} holds for dipoles near arbitrary gain-loss resonators, it must also hold for a system consisting of a dipole kept near a time-varying sphere. However, Equation~\eqref{eq:balance} was initially written for dipoles near static gain media \cite{ren2024classical}. Therefore, it must be first verified for the system involving time-varying media. To test Equation~\eqref{eq:balance}, we plot the values of the left and right-hand sides independently as a function of the emission frequency $\omega_\mathrm{d}$ of the dipole in Figure~\ref{fig:power_balance}(b). In the context of such a sphere-dipole system, $P_\mathrm{far}$, $P_\mathrm{LDOS}$, and $P_\mathrm{gain}-P_\mathrm{loss}$ refer to the net powers crossing the surfaces $A_\mathrm{f}$, $A_\mathrm{d}$, and $A_\mathrm{s}$, respectively (see Figure~\ref{fig:power_balance}(a)). Here, $A_\mathrm{f}$, $A_\mathrm{d}$, and $A_\mathrm{s}$ correspond to the surfaces enclosing both the sphere and dipole, only the dipole, and only the sphere, respectively. Note that we use the T-matrix method to calculate these powers (see Supporting Information; Section~S4). Furthermore, in Figure~\ref{fig:power_balance}(b), the plotted powers are normalized by the power $P_0$ radiated by the dipole in a vacuum. From Figure~\ref{fig:power_balance}(b), we observe that the power conservation relation given in Equation~\eqref{eq:balance} also holds for the sphere-dipole system. Next, the gain corrected expression for the total emitted power by the dipole $P^\mathrm{dip}_\mathrm{total}$ that is responsible for its spontaneous decay in the presence of such a non-Hermitian sphere is \cite[Equation~(23)]{ren2024classical}
\begin{eqnarray}
P^\mathrm{dip}_\mathrm{total}= P_\mathrm{LDOS}+P_\mathrm{gain}\,.
\label{eq:total}
\end{eqnarray}
Moreover, using Equation~\eqref{eq:balance} and \eqref{eq:total}, we can write $P_\mathrm{far}=P^\mathrm{dip}_\mathrm{total}-P_\mathrm{abs}$. Note that $P^\mathrm{dip}_\mathrm{total}-P_\mathrm{abs}$ corresponds to the net radiatively emitted power from the dipole $P^\mathrm{dip}_\mathrm{rad}$ that leads to its spontaneous \textit{radiative} decay \cite{carminati2006radiative,moroz2010nonradiative}. Therefore, we conclude that $P_\mathrm{far}=P^\mathrm{dip}_\mathrm{rad}$. Finally, the radiative decay rate enhancement ${\gamma_\mathrm{rad}}/{\gamma_\mathrm{0}}$ of the dipole kept near a time-varying sphere can be written as 
\begin{eqnarray}
\frac{\gamma_\mathrm{rad}}{\gamma_\mathrm{0}}=\frac{P^\mathrm{dip}_\mathrm{rad}}{P_\mathrm{0}}=\frac{P_\mathrm{far}}{P_\mathrm{0}}\,.
\label{eq:enhancement}
\end{eqnarray}
Here, $\gamma_\mathrm{rad}$ refers to the radiative decay rate of the dipole in the presence of the time-varying sphere, and $\gamma_\mathrm{0}$ refers to the decay rate of the dipole in a vacuum. 

Having discussed the details of decay rate calculation, we proceed to apply our inverse design framework to engineer the radiative decay rate enhancement of the dipoles near time-varying spheres on demand.

\subsection{Anomalous Drexhage Effect}
In 1970, K.H. Drexhage showed that the radiative decay rate enhancement of a dipole kept near a perfect mirror depends on the orientation of the dipole to the surface of the mirror \cite{Drexhage1970influence}. A similar effect is observed for a dipole near a \textit{static} sphere, as shown in \textbf{Figure~\ref{fig:drexhage}}(a). Here, we plot the radiative decay rate enhancement ${\gamma_\mathrm{rad}}/{\gamma_\mathrm{0}}$ of an electric dipole for its two orthogonal polarizations as a function of the distance ${r}_\mathrm{d}$ from the center of a static sphere. Note that $\perp$-polarization refers to the radial and $\parallel$-polarization refers to the tangential orientation of the dipole with respect to the sphere. The radius of the sphere is $R=10$~nm, and the emission frequency of the dipole is $\omega_\mathrm{d}=0.072c_0/R$. Here, $c_0$ is the speed of light in a vacuum.

From Figure~\ref{fig:drexhage}(a), we observe that the decay rate enhancement shows different trends as a function of the distance ${r}_\mathrm{d}$ for the two orientations of the dipole. Consider the regime $r_\mathrm{d}/R\approx1$. For the tangential orientation, the decay rate is almost negligible leading to ${\gamma_\mathrm{rad}}/{\gamma_\mathrm{0}}\approx0$. In contrast, the decay rate is significant for the radial orientation leading to ${\gamma_\mathrm{rad}}/{\gamma_\mathrm{0}}\gg1$. Such an orientation-dependent decay rate enhancement of the dipole is known as the Drexhage effect \cite{Drexhage1970influence}. 

The physical explanation for the Drexhage effect is as follows. To begin with, the decay rate enhancement of the dipole depends on the total far field \cite{Krasnok2015antenna,carminati2006radiative} (see Equation~\eqref{eq:enhancement}). The total far field $\mathbf{E}^\mathrm{tot}(\mathbf{r}_\mathrm{far})$ is the sum of the field scattered from the sphere $\mathbf{E}^\mathrm{sca}(\mathbf{r}_\mathrm{far})$ and the field of the dipole itself $\mathbf{E}^\mathrm{dip}(\mathbf{r}_\mathrm{far})$, in the absence of the sphere, at a spatial location $\mathbf{r}=\mathbf{r}_\mathrm{far}$. Here, $r_\mathrm{far}\gg R$. For the example shown in Figure~\ref{fig:drexhage}(a), $\mathbf{E}^\mathrm{sca}$ and $\mathbf{E}^\mathrm{dip}$ interfere destructively (constructively) for the tangential (radial) orientation of the dipole leading to negligible (significant) decay rate enhancement. Besides, for $r_\mathrm{d}/R\gg1$, for both the orientations of the dipole, ${\gamma_\mathrm{rad}}/{\gamma_\mathrm{0}}$ saturates to unity. This saturation happens because $\mathbf{E}^\mathrm{sca}$ decays rapidly as $r_\mathrm{d}$ increases, leading to a decrease in the interference effects of the fields and consequently the impact of the sphere on $\gamma_\mathrm{rad}$ \cite{carminati2006radiative}. Therefore, $\gamma_\mathrm{rad}$ approaches $\gamma_\mathrm{0}$ for $r_\mathrm{d}/R\gg1$.

Having discussed the Drexhage effect, we explore the anomalous Drexhage effect next. It refers to a situation where a dipole shows a similar decay rate enhancement for its two orthogonal orientations. This is considered here as anomalous because it is in striking contrast to the Drexhage effect that has been studied from multiple perspectives \cite{lutz2016drexhage,kwandrin2012graytone}. Our goal is to show such an anomalous Drexhage effect by an appropriate time modulation of the sphere. To do so, we design the time-varying electron density $N(t)$ of the sphere such that, even for the tangential orientation of the dipole, we have a significant decay rate enhancement for $r_\mathrm{d}/R\approx1$. Therefore, we maximize $\gamma_\mathrm{rad}/\gamma_0$ for $r_\mathrm{d}/R\approx1$ for the tangential orientation of the dipole. 
From Equation~\eqref{eq:enhancement}, we observe that $\gamma_\mathrm{rad}/\gamma_0$ is proportional to the power emitted in the far field by the sphere-dipole system ${P_\mathrm{far}}$. Consequently, our task at hand is reduced to maximizing ${P_\mathrm{far}}$ when the dipole is placed sufficiently close to the sphere. We choose $r_\mathrm{d}=1.01R$. 

To set up the optimization problem, we define our objective function $f_\mathrm{obj}$ in terms of ${P_\mathrm{far}}$ (see the Methods section). Note that besides ${P_\mathrm{far}}$, $f_\mathrm{obj}$ also consists of a penalty function that penalizes the optimizer if $N(t)$ tends toward negative values for any $t \in [0,\tau_\mathrm{m}]$. We use such a penalty function with all other objective functions in this article. Additionally, we need to fix the base modulation frequency $\omega_\mathrm{m}$ \textit{a priori} in the optimization. To choose $\omega_\mathrm{m}$ such that it assists the maximization of $P_\mathrm{far}$, we first plot the amplitude of the electric dipolar Mie coefficient $a_\mathrm{ED}$ of the static sphere as a function of the excitation frequency $\omega$ \cite{Rahimzadegan2020minimalist} (see Figure~\ref{fig:drexhage}(b)). From Figure~\ref{fig:drexhage}(b), we observe the existence of the electric dipolar resonance of the sphere at $\omega=\omega_\mathrm{ED}$. Therefore, we choose $\omega_\mathrm{m}=\omega_\mathrm{ED}-\omega_\mathrm{d}$. Such a choice is made because once the time modulation is turned on, the sphere in the presence of the dipole emitting at the frequency $\omega_\mathrm{d}$ scatters the frequencies $\omega_j=\omega_\mathrm{d}+j\omega_\mathrm{m}$, with $j\in\mathbb{Z}$ (see Figure~\ref{fig:concept}(a)). Therefore, $\omega_\mathrm{1}=\omega_\mathrm{ED}$ lies exactly at the electric dipolar resonance frequency leading to a resonant enhancement of ${P_\mathrm{far}}$ in the time-varying case for an appropriately optimized $N(t)$ (see Equation~(S6) of Supporting Information).

Having chosen $f_\mathrm{obj}$ and $\omega_\mathrm{m}$, we use the inverse design framework discussed in Section~\ref{sec:inverse} to maximize $\gamma_\mathrm{rad}/\gamma_0$ for the dipole oriented tangentially to the time-varying sphere at $r_\mathrm{d}=1.01R$. The optimized $N(t)$ is shown in Figure~\ref{fig:drexhage}(c). Note that within our optimization landscape, there exist different temporal profiles that provide similar results. Ultimately, our goal is to obtain a design that fulfills our objective. Thus, any such solution is suitable. This conclusion also holds for all other optimization results presented in this work. Furthermore, in Figure~\ref{fig:drexhage}(d), we plot $\gamma_\mathrm{rad}/\gamma_\mathrm{0}$ as a function of $r_\mathrm{d}$ for the dipole near the time-varying sphere. From Figure~\ref{fig:drexhage}(d), we observe that for both the tangential and radial orientation of the dipole, the decay rate enhancement is significant (i.e., $\gamma_\mathrm{rad}/\gamma_\mathrm{0}\gg1$) for $r_\mathrm{d}/R\approx1$, leading to a demonstration of the anomalous Drexhage effect. 

In particular, as intended, the optimization successfully returns an $N(t)$ that maximizes $\gamma_\mathrm{rad}/\gamma_\mathrm{0}$ when $r_\mathrm{d}/R\approx1$ for the tangential orientation of the dipole. Such a maximization can be explained as follows. As expected, the temporal modulation with the optimized $N(t)$ enhances the total far field $\mathbf{E}^\mathrm{tot}(\mathbf{r}_\mathrm{far})$ at $\omega_j=\omega_1$ (see Supplementary Information; Section~S4). Moreover, since the power $P_\mathrm{far}$ depends on the contributions of $\mathbf{E}^\mathrm{tot}(\mathbf{r}_\mathrm{far})$ across all the frequencies $\omega_j$, this field enhancement at $\omega_1$ maximizes $P_\mathrm{far}$, which in turn maximizes the decay rate enhancement. Additionally, even for the radial orientation of the dipole, $\gamma_\mathrm{rad}/\gamma_\mathrm{0}$ at $r_\mathrm{d}/R\approx1$ is about $300$ times higher in the time-varying case than that in the static case. Such a high  $\gamma_\mathrm{rad}/\gamma_\mathrm{0}$ for the radial orientation also occurs due to a similar total far field enhancement.

It is worth remarking that in order to instead suppress $\gamma_\mathrm{rad}$ for the radial orientation of the dipole shown in Figure~\ref{fig:drexhage}(a), one needs to reduce $P_\mathrm{far}$ sufficiently (see Equation~\eqref{eq:enhancement}). One possible way to achieve this is to enhance the absorptivity of the sphere using time modulation \cite{hayran2024beyond,mostafa2022coherently}. In the limiting case when the sphere acts as a perfect absorber, $P_\mathrm{far}$ vanishes, leading to a complete suppression of $\gamma_\mathrm{rad}$.

As discussed in the following, we now apply our inverse design framework to a system involving spatiotemporal metasurfaces.
\begin{figure*}
\centerline{\includegraphics[width= 0.8\columnwidth,trim=1 1 0.5 1,clip]{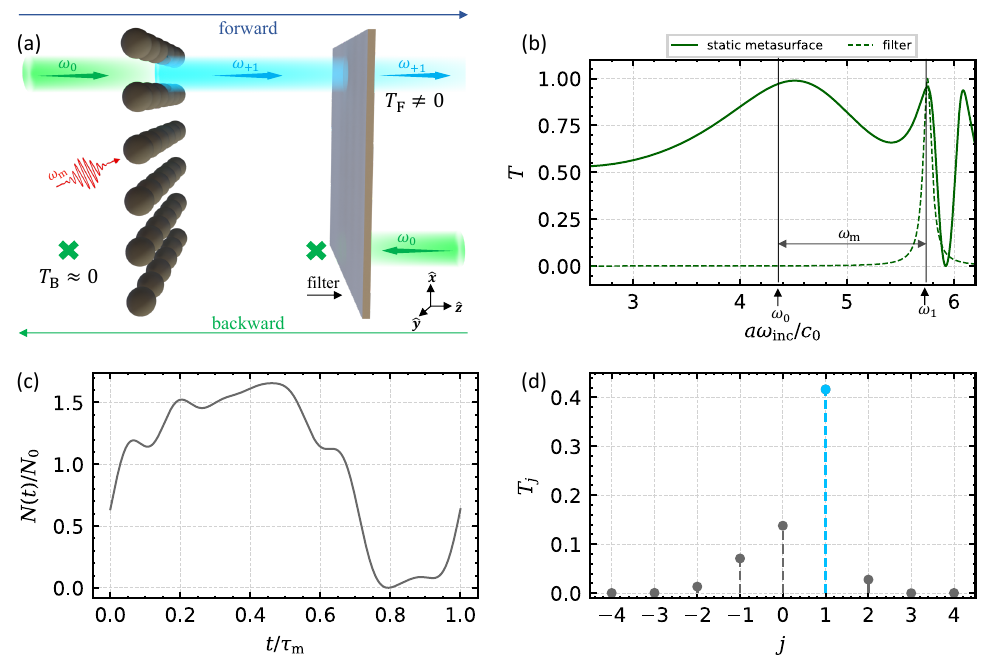}}
\caption{Asymmetric transmission (AT) of light at visible frequencies. (a) A conceptual illustration of an AT system formed by cascading a spatiotemporal metasurface and a narrow-band filter. Here, $T_\mathrm{F}$ ($T_\mathrm{B}$) corresponds to the final forward (backward) transmissivity of the AT system. (b) The transmissivity $T$ of the metasurface under static operation (solid curve) and filter (dotted curve) as shown in (a) as a function of the incident frequency $\omega_\mathrm{inc}$. (c), (d) The optimized electron density $N(t)$ as a function of time $t$, and the corresponding transmissivity contribution $T_j$ of the spatiotemporal metasurface as a function of the harmonic number $j$, respectively.}
\label{fig:asymmetric}
\end{figure*}
\section{Asymmetric transmission of light at visible frequencies}

As the second application of our inverse design framework, we design a device that implements asymmetric transmission (AT) of light at visible frequencies. A device operating under the AT condition allows efficient transmission of light in one direction while blocking it in the opposite direction \cite{ansari2023asymmetric}. A conceptual illustration of spatiotemporal metasurface-based AT is shown in \textbf{Figure~\ref{fig:asymmetric}}(a). The system shown in Figure~\ref{fig:asymmetric}(a) is formed by cascading a spatiotemporal metasurface and a narrow-band frequency filter. 

When an electromagnetic wave at frequency $\omega_0$ is incident to the system from the left-hand side (forward propagation), it encounters the spatiotemporal metasurface that upconverts most of its energy to an electromagnetic wave at frequency $\omega_1=\omega_0+\omega_\mathrm{m}$. Our design objective is to render this upconversion efficient. The wave then encounters the filter that offers complete transmission to fields with frequency $\omega_1$ while blocking all other frequencies \cite{niraula2015single,Tibuleac1997reflection}. Therefore, the wave at frequency $\omega_1$ gets completely transmitted through the filter, leading to a non-zero final forward transmissivity $T_\mathrm{F}$. On the other hand, if a wave at frequency $\omega_0$ is incident from the right-hand side to the system (backward propagation), it encounters the filter first, which does not allow any further propagation of the wave. Therefore, the final backward transmissivity $T_\mathrm{B}$ of the whole system is negligible. Moreover, it is important to mention that the problem of frequency upconversion using spatiotemporal structures was also studied in various other works \cite{liu2018huygens,wu2020serrodyne,sedeh2022optical,ramaccia2020electromagnetic,koutserimpas2018nonreciprocal,sisler2024electrically,zongfu2009complete}. However, all these works were limited to the adiabatic regime of temporal modulation (i.e., $\omega_\mathrm{m}\ll\omega_0$), which requires the use of frequency filters with very narrow bandwidths. Furthermore, the frequency of operation in these works was either in the microwave or infrared regime. In our framework, we use non-adiabatic temporal modulation, our metasurface operates at visible frequencies, and we fully take into account the effects of material dispersion that cannot be ignored for most of the materials at such frequencies.

We aim to design an AT system as shown in Figure~\ref{fig:asymmetric}(a). A critical component of the AT system is the spatiotemporal metasurface, which is capable of frequency upconversion. In what follows, we discuss the choice of parameters of the system deemed to be supportive in achieving an efficient frequency upconversion using the Mie resonances of the underlying static metasurface \cite{rahimzadegan2022comprehensive,panagiotidis2023optical}. To begin with, we assume the metasurface to be made from a square lattice of spheres. The radius $R$ of the spheres is $R=150$~nm, and the lattice constant $a$ of the metasurface is $a=2.6R$. Besides, for simplicity, the parameters are chosen such that the metasurface is subwavelength in the spectral range of interest. For a subwavelength metasurface, only the principle spatial diffraction order is propagating. The transmissivity $T$ of the metasurface as a function of the incident frequency $\omega_\mathrm{inc}$ under static conditions (i.e., $N(t)=N_0$) is shown in Figure~\ref{fig:asymmetric}(b). Note that here, and for the following simulations, we assume a monochromatic $x$-polarized plane wave normally incident on the metasurface for calculating $T$. From Figure~\ref{fig:asymmetric}(b), we observe that the static metasurface supports various Mie resonances \cite{rahimzadegan2022comprehensive}. 

Using the transmission spectrum of the static metasurface, we choose the frequencies $\omega_0$ and $\omega_\mathrm{m}$ for the time-varying operation (see Figure~\ref{fig:asymmetric}(b)). From Figure~\ref{fig:asymmetric}(b), it is apparent that $\omega_0$ and $\omega_\mathrm{m}$ are chosen such that $\omega_1$ lies at a high-quality resonance of the static metasurface. Such a choice is made so that upon temporal modulation of the metasurface with an appropriately chosen $N(t)$, a resonant enhancement of the transmitted field occurs at $\omega_1$ for an incident field at $\omega_0$. Importantly, to design the AT functionality of the spatiotemporal metasurface for visible frequencies, the parameters $a,\,R$, and $\omega_\mathrm{m}$ are chosen such that $\omega_0$ and $\omega_1$ lie in the visible spectrum. In our example, $\omega_0=2\pi \times 542$~THz and  $\omega_1=2\pi \times 704$~THz. 

Next, we use our inverse design framework to optimize the time-varying electron density $N(t)$ of the spatiotemporal metasurface. The total transmissivity $T$ of the spatiotemporal metasurface can be written as a sum of the contribution of all the transmitted frequency harmonics, i.e., $T=\sum_jT_j$ \cite{garg2022modeling} (see Figure~\ref{fig:concept}(a)). Our goal is to maximize the transmissivity at $\omega_1$. Therefore, we choose the objective function $f_\mathrm{obj}$ in terms of $T_1$ (see the Methods section). 

The electron density $N(t)$ as a result of the optimization is shown in Figure~\ref{fig:asymmetric}(c). Furthermore, the corresponding transmissivity contributions $T_j$ as a function of the scattered harmonic number $j$ are shown in Figure~\ref{fig:asymmetric}(d). From Figure~\ref{fig:asymmetric}(d), we observe that the metasurface with the optimal $N(t)$ preferentially transmits the energy of the incident wave to the frequency $\omega_1$ (with $T_1=0.42$). Such a preferential transmission can be explained as follows. To begin with, the time-varying electron density $N(t)$ effectively modulates the amplitude and phase of the time-domain complex transmission coefficient of the metasurface \cite{liu2018huygens} (see Supporting Information; Section~S7). In principle, such a modulation of the transmission coefficient gives rise to the transmitted frequencies $\omega_j$ (with $j\in\mathbb{Z}$). However, due to the particular choice of $f_\mathrm{obj}$, the optimizer converges such that this amplitude and phase modulation leads to an efficient coupling to $\omega_1$ in transmission (also see Supporting Information; Section~S8). Additionally, the high-quality resonance of the underlying static metasurface at $\omega_1$ assists in the resonant enhancement of such a coupling (see Figure~\ref{fig:asymmetric}(b)). It is important to remark that such a preferential transmission is insensitive to the phase of the incident wave.

Note that the total transmission efficiency of the optimized spatiotemporal metasurface is about $66\%$. Approximately $15\%$ of the incident energy is lost to reflection and about $19\%$ to material losses due to frequency dispersion. Moreover, from Figure~\ref{fig:asymmetric}(d), we observe that even though the maximum transmitted energy is linked to $\omega_1$, a non-zero amount of energy is coupled to other scattered frequencies. To attain a perfect upconversion to the frequency $\omega_1$ (i.e., $T_1=1$ and $T_j=0\,\forall\,j\neq1$), one needs a spatiotemporal metasurface whose time-domain complex transmission coefficient has a pure linear phase modulation of $2\pi$ in the time-period $\tau_\mathrm{m}$ \cite{liu2018huygens} (see Supporting Information; Section~S7). Attaining such an idealized condition is challenging at the optical frequencies \cite{sedeh2020adaptive}. 

Next, we choose a filter that has almost unity transmissivity $T$ at $\omega_1$ and negligible $T$ at all other frequencies (see the dotted curve in Figure~\ref{fig:asymmetric}(b)). We implement the filter by optimizing a Fabry-Perot cavity (see Supporting Information; Section~S9). Other possible implementations of such a filter can be found in \cite{niraula2015single,Tibuleac1997reflection}. Finally, cascading such a filter with the optimized metasurface results in a final forward transmissivity of $T_\mathrm{F}=0.41562$ for a wave incident at $\omega_0=2\pi \times 542$~THz. Furthermore, the final backward transmissivity of the system for the same incident wave is $T_\mathrm{B}=0.00097$. Therefore, our inverse design framework allows us to realize the asymmetric transmission of light at visible frequencies. 

It is important to mention that this AT is different from the phenomena of nonreciprocity \cite{asadchy2020tutorial}. We present a test of nonreciprocity for the cascaded in the Supporting Information in Section~S10. We perform this test by considering a backward propagating incident field to the filter at frequency $\omega_1$. We find that since the electron density profile $N(t)$ in Figure~\ref{fig:asymmetric}(c) breaks the generalized time-reversal symmetry, the cascaded system is nonreciprocal \cite{williamson2020integrated}. Note that another trivial $N(t)$ profile known in the literature that leads to the breaking of the time-reversal symmetry and hence reciprocity is a sawtooth-type modulation \cite{williamson2020integrated,liu2018huygens,wu2020serrodyne}.

\section{Conclusion}
We have presented a differentiable T-matrix-based inverse design framework to engineer the scattering response of time-varying spheres and spatiotemporal metasurfaces on demand. These time-varying nanostructures are assumed to be made from a dispersive material with a time-varying electron density. Using our inverse design framework, we optimized the temporal profile of the electron density to tailor the desired functionality of the chosen nanostructure. We used a gradient-based approach to perform the optimization. We exploited our inverse design framework in two specific examples.

First, we computed the decay rate enhancement of dipoles kept near time-varying spheres. In particular, we optimized time-varying spheres such that the dipoles exhibit anomalous Drexhage effect. Next, we applied our inverse design framework to spatiotemporal metasurfaces. We optimized the metasurface to achieve an efficient frequency upconversion. Such an optimized metasurface is then cascaded with a narrow-band filter so that the composite system supports asymmetric transmission (AT) of light at visible frequencies. This AT operation is achieved while using non-adiabatic temporal modulation and considering the effects of material dispersion.

The T-matrix-based inverse design tool introduced in this article assists in simultaneously controlling the spatial and temporal properties of light. Depending on the choice of the objective function, various applications using the time-varying nanostructures can be realized in principle. From this perspective, our approach unlocks the opportunities to design novel photonic time and space-time crystals.


\section{Methods}
\textit{Further details on the differentiable T-matrix method}: Let us assume a general polychromatic incident excitation to the considered time-varying nanostructure with frequencies, $\omega_j=\omega+j\omega_\mathrm{m}$ with $j\in[-J,J]$. Here, $J$ is an integer that should be chosen sufficiently large to ensure numerical convergence. Then, the incident field ${\mathbf{E}}^{\mathrm{inc}}$ and scattered field ${\mathbf{E}}^{\mathrm{sca}}$ can be written as
\begin{subequations}\label{eq:inc_sca}
\begin{align}
    {\mathbf{E}}^{\mathrm{inc}}(\mathbf{r},t)&=\sum_{jl\mu s}A_{jl\mu s}^{\mathrm{inc}}\mathbf{F}^{(1)}_{j\mu s}(k_j\mathbf{r})\mathrm{e}^{-i\omega_j t}\,,\label{eq:E_inc}\\
    {\mathbf{E}}^{\mathrm{sca}}(\mathbf{r},t)&=\sum_{jl\mu s}A_{jl\mu s}^{\mathrm{sca}}\mathbf{F}^{(3)}_{l\mu s}(k_j\mathbf{r})\mathrm{e}^{-i\omega_jt}\,,\label{eq:E_sca}
    \end{align}
    \end{subequations}
    \noindent
where $k_j= \omega_j/c_0$. Besides, $\mathbf{F}^{(1)}_{l\mu s}(x)$ ($\mathbf{F}^{(3)}_{l\mu s}(x)$) represent the regular (radiating) vector spherical harmonics (VSHs) with total angular momentum $l=1,2,3...,l_\mathrm{max}$, $z$-component of angular momentum $\mu= -l, -l+1,...,l$, and parity $s= 0,1$. Here, $s=0$ represents the transverse-electric (TE), and $s=1$ represents the transverse-magnetic (TM) mode, respectively. Moreover, $l_\mathrm{max}$ is the maximum multipolar order retained in the expansion. It should be chosen sufficiently large to ensure numerical convergence. Next, using the method in \cite{ptitcyn2023floquet,garg2022modeling}, one can connect $\mathbf{A}^\mathrm{inc}$ and $\mathbf{A}^\mathrm{sca}$ by the T-matrix $\mathbf{\hat{T}}$ as given in Equation~\eqref{eq:Tmat}. Note that $\mathbf{\hat{T}}$ is a square matrix with size $2l_\mathrm{max}(2J+1)(l_\mathrm{max}+2)$. 

\noindent\textit{The objective function for maximizing $\gamma_\mathrm{rad}/\gamma_\mathrm{0}$}: From Equation~\eqref{eq:enhancement}, we observe that $\gamma_\mathrm{rad}$ is proportional to $P_\mathrm{far}$. Therefore, maximizing $\gamma_\mathrm{rad}$ requires the maximization of $P_\mathrm{far}$. For the simulation result shown in Figure~\ref{fig:drexhage}(c) and (d), we found the minimization of the objective function to perform better in terms of convergence. Therefore, we defined the objective function to be minimized as $f_\mathrm{obj}(\mathbf{M})=w/P_\mathrm{far}(\mathbf{M})+(1-w)p(\mathbf{M})$. Here, $p(\mathbf{M})$ is a penalty function that penalizes the optimizer if it goes toward those values of $\mathbf{M}$ that correspond to a negative $N(t)$ for any $t\in [0,\tau_\mathrm{m}]$ (see Supporting Information; Subsection~3.2 for more details on $p(\mathbf{M})$). Furthermore, $w$ is a quantity that distributes the weight of the objective function between $1/P_\mathrm{far}$ and the penalty $p$. We chose $w=0.99$ for this optimization. 

\noindent\textit{The objective function for maximizing $T_1$}: For the example shown in Figure~\ref{fig:asymmetric}(a), we aim to maximize $T_1$. Therefore, we define the objective function to be maximized as $f_\mathrm{obj}(\mathbf{M})=wT_1(\mathbf{M})-(1-w)p(\mathbf{M})$. Here, as we are maximizing $f_\mathrm{obj}$, $p(\mathbf{M})$ has a negative weight. We chose $w=0.84$ for this optimization.

\medskip
\noindent\textbf{Supporting Information} \par
\noindent Supporting Information is available from the Wiley Online Library or from the author.

\medskip
\noindent\textbf{Acknowledgments} \par
The authors would like to thank Dr. Elli Stamatopoulou, and Dr. Markus Nyman for useful discussions about the decay rate enhancement. P.G. and C.R. are part of the Max Planck School of Photonics, supported by the Bundesministerium für Bildung und Forschung, the Max Planck Society, and the Fraunhofer Society. P.G. acknowledges support from the Karlsruhe School of Optics and Photonics (KSOP). P.G. and C.R. acknowledge financial support by the German Research Foundation within the SFB 1173 (Project-ID No. 258734477). J.D.F. and C.R. acknowledge financial support by the Helmholtz Association in the framework of the innovation platform “Solar TAP”. C.R. acknowledges support from the German Research Foundation within the Excellence Cluster 3D Matter Made to Order (EXC 2082/1 under project number 390761711) and by the Carl Zeiss Foundation.
T.J.S. acknowledges funding from the Alexander von Humboldt Foundation.~M.S.M. acknowledges support from the Research Council of Finland (Grant No.~336119).
V.A. acknowledges the Research Council of Finland (Project No. 356797), Finnish Foundation for Technology Promotion, and  Research Council of Finland Flagship Programme, Photonics Research and Innovation (PREIN), decision number 346529, Aalto University. X.W. acknowledges the Fundamental Research Funds for the Central Universities, China (Project No. 3072024WD2603) and the Scientific Research Foundation, Harbin Engineering University, China (Project No. 0165400209002).

\medskip
\bibliographystyle{MSP}
\bibliography{references}

\section*{Supplementary Material}
\setcounter{section}{0}
\renewcommand{\thesection}{S\arabic{section}}
\setcounter{figure}{0}
\renewcommand{\thefigure}{S\arabic{figure}}
\setcounter{equation}{0}
\renewcommand{\theequation}{S\arabic{equation}}

\section{Drude model with time-varying electron density}

In the main text, we consider nanostructures made from time-varying dispersive media. To model the dispersion of the media, we use the Drude model. The current density $\mathbf{J}(\mathbf{r},t)$ of the dispersive media with time-varying electron density $N(t)$ according to the Drude model satisfies \cite{ptitcyn2023floquet,garg2022modeling}

\begin{eqnarray}
\left[\frac{\partial}{\partial t}+\eta\right]\mathbf{J}(\mathbf{r},t)=\frac{N(t)e^2}{m_\mathrm{e}}\mathbf{E}(\mathbf{r},t)\,.\label{drude}
    \end{eqnarray}
    \noindent
Here, $e$ and $m_\mathrm{e}$ correspond to the charge and mass of an electron, respectively. Further, $\eta$ is the damping factor, and $\mathbf{E}(\mathbf{r},t)$ is the electric field that drives the electron in the Drude model. For the simulations in the manuscript, we assume the spheres to be made from gold. Therefore, we use $\eta=27.91$~THz and $N_0=5.30\times10^{28}$~$\mathrm{m}^{-3}$ \cite{blaber2009search}.

\section{Relation between the T-matrix and S-matrix formalism}
For an isolated sphere with T-matrix $\mathbf{\hat{T}}_\mathrm{s}$, its S-matrix can be calculated as \cite{waterman2009tmatrix}
\begin{equation}
\mathbf{\hat{S}}_\mathrm{s}= \left(\mathbf{\hat{I}}+2\mathbf{\hat{T}}_\mathrm{s}\right)\,.\label{eq:Smatsphere}
\end{equation}

Further, for a metasurface made from such spheres, we first write its effective T-matrix $\mathbf{\hat{T}}_\mathrm{eff}$ as 

\begin{equation}
\mathbf{\hat{T}}_\mathrm{eff}= \left(\mathbf{\hat{I}}-\mathbf{\hat{T}}_\mathrm{s}\mathbf{\hat{C}}\right)^{-1}\mathbf{\hat{T}}_\mathrm{s}\,.\label{eq:Teff}
\end{equation}
Here, $\mathbf{\hat{I}}$ is the identity matrix, and $\mathbf{\hat{C}}$ is a matrix that captures the coupling of the spheres in the lattice. From Equation~\eqref{eq:Teff}, we note that once the T-matrix $\mathbf{\hat{T}}_\mathrm{s}$ of an isolated sphere is known, we only need $\mathbf{\hat{C}}$ to compute the effective T-matrix of the metasurface. Furthermore, $\mathbf{\hat{C}}$ can be efficiently calculated using the Ewald summation method as shown in \cite{ewald1921die,beutel2021efficient,beutel2024treams}. Finally, one can compute the S-matrix for the metasurface once $\mathbf{\hat{T}}_\mathrm{eff}$ is known by using
\begin{equation}
\label{eq:Teff_c}
\mathbf{\hat{S}}=\mathbf{\hat{I}}+\mathbf{\hat{V}}\hspace{2pt}\mathbf{\hat{T}_\mathrm{eff}}\hspace{2pt}\mathbf{\hat{U}}\,,
\end{equation}
\noindent
where $\mathbf{\hat{U}}$ is the coordinate transformation matrix of the incident field from the plane to the spherical wave basis, and $\mathbf{\hat{V}}$ is the coordinate transformation matrix of the scattered field from the spherical to plane wave basis \cite[Eqs.~(17)-(18)]{garg2022modeling}.

\section{Gradient-based optimization}
\subsection{Details of the used software packages}
As mentioned in the main text, we use gradient-based optimization to calculate the optimal Fourier coefficients of $N(t)$ that minimize/maximize the objective function $f_\mathrm{obj}$. Such a gradient-based optimization approach requires us to compute the derivatives of $f_\mathrm{obj}$ with respect to $\mathbf{M}$. Therefore, we write our code using the software package \colorbox{mygray}{\texttt{jax}} as it can automatically differentiate native Python and Numpy functions using the \colorbox{mygray}{\texttt{jax.grad}} utility \cite{jax2018github}. Further, to perform the optimization using the gradient information, we use the software package \colorbox{mygray}{\texttt{nlopt}} \cite{NLopt}. In particular, we use the method \colorbox{mygray}{\texttt{NLOPT\_\,LD\_\,MMA}} which is an implementation of globally convergent method-of-moving-asymptotes \cite{svanbergaclass2002}.

\subsection{Details of the penalty function}

\begin{figure*}
\centerline{\includegraphics[width= 0.5\columnwidth]{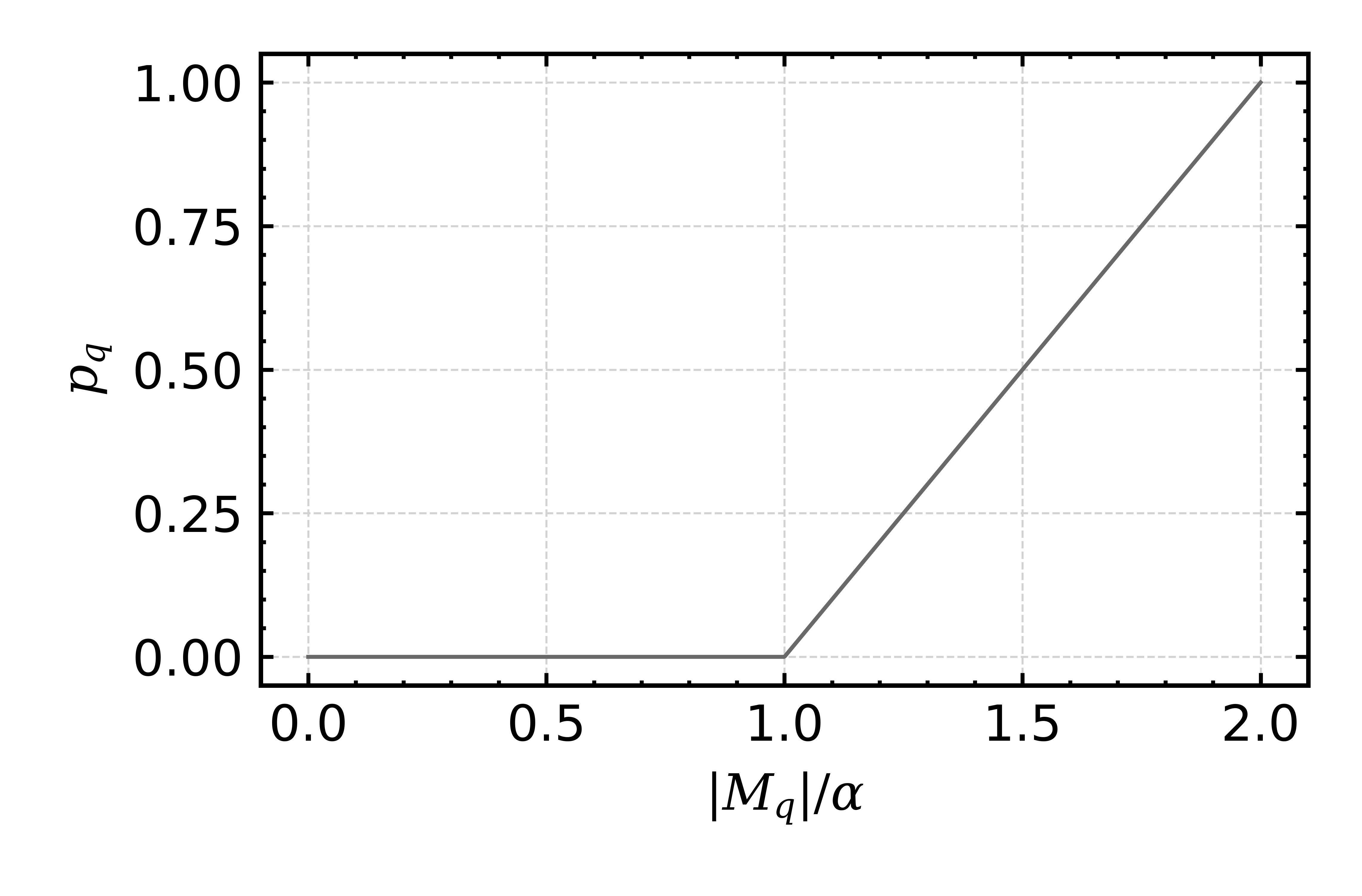}}
\caption{Variation of the penalty contribution $p_{q}$ as a function of the absolute value of the $q$-th Fourier coefficient $M_q$. The penalty function $p$ is defined as $p=\sum_{q=1}^{Q}p_q$.}
\label{fig:penalty}
\end{figure*}

Further, as discussed in the main text, we use a penalty function $p(\mathbf{M})$ to steer the optimizer away from those values of $\mathbf{M}$ that lead to a negative $N(t)$ for any $t\in[0,\tau_\mathrm{m}]$. As ${N(t)=N_0\left[1+\sum_{q=1}^{Q}(M_q e^{-iq\omega_\mathrm{m}t}+\textrm{c.c.})\right]}$, one way to define $p$ is such that the optimizer gets penalized if any $|M_q|$ crosses a certain threshold $\alpha$. This is because the absolute values of the Fourier coefficients $M_{q}$ quantify the modulation amplitude of $N(t)$. Therefore, if any $|M_q|$ crosses a certain threshold, $N(t)$ might become negative for some $t \in [0,\tau_\mathrm{m}]$. Consequently, we define $p(\mathbf{M})=\sum_{q=1}^{Q}p_q(\mathbf{M})$ with $p_q(\mathbf{M})=\textrm{max}(|M_q|,\alpha)-\alpha$. Here, if $|M_q|\leq\alpha$, then $p_q=0$, else $p_q>0$. Further, $p_q$ increases linearly with $|M_q|$ if $|M_q|>\alpha$ (see Figure~\ref{fig:penalty}). 

\section{Field and power calculation for oscillating dipoles}
One can also incorporate the fields of oscillating dipoles in the T-matrix method. Consider an oscillating point dipole placed at a spatial location $\mathbf{r}_\mathrm{d}$ emitting a monochromatic field at a frequency $\omega_\mathrm{d}$. Classically, such a dipole can be modeled by a current density given as $\mathbf{J}_\mathrm{d}(\mathbf{r},t)=-i\omega_\mathrm{d}\mathbf{p}_0\delta(\mathbf{r}-\mathbf{r}_\mathrm{d})e^{-i\omega_\mathrm{d}t}$. Here, $\mathbf{p}_0$ is the dipole moment of the dipole and $\delta$ is the Dirac delta distribution. The electric field $\mathbf{E}^\mathrm{dip}$ emitted by such a dipole can be expanded in terms of the regular and radiating vector spherical harmonics (VSHs) as

\begin{subequations}\label{eq:dipole_field}
\begin{align}
    {\mathbf{E}}^{\mathrm{dip}}(\mathbf{r},t)&=\sum_{jl\mu s}d_{jl\mu s}^{(1)}\mathbf{F}^{(1)}_{l\mu s}(k_j\mathbf{r})\mathrm{e}^{-i\omega_j t}\hspace{10pt}\textrm{and}\label{eq:dip_regular}\\
    {\mathbf{E}}^{\mathrm{dip}}(\mathbf{r},t)&=\sum_{jl\mu s}d_{jl\mu s}^{(3)}\mathbf{F}^{(3)}_{l\mu s}(k_j\mathbf{r})\mathrm{e}^{-i\omega_j t}\,\textrm{, respectively}.\label{eq:dip_radiating}
    \end{align}
    \end{subequations}
    \noindent
Here, $\omega_0=\omega_\mathrm{d}$. Note that for a monochromatic dipole radiating at the frequency  $\omega_0=\omega_\mathrm{d}$, $d_{jl\mu s}^{(1)/(3)}=0\, \forall j\neq 0$. Further, because the origin of the coordinate system is not necessarily at the location of the dipole, $d_{jl\mu s}^{(3)}$ can be non-zero for $l\neq1$. We use the software package \colorbox{mygray}{\texttt{treams}} to calculate $d_{jl\mu s}^{(1)/(3)}$ \cite{beutel2024treams}. In what follows, we use the field information of the dipoles to calculate the power flows through various surfaces of interest.

As mentioned in the main text, a system consisting of a dipole placed near a time-varying sphere satisfies the power conservation relation, $P_\mathrm{far}= P_\mathrm{LDOS}+P_\mathrm{gain}-P_\mathrm{loss}$ (see Figure~2). Here, we show the explicit expression of these powers.

To calculate the power radiated by the sphere-dipole system in the far field $P_\mathrm{far}$, one needs to integrate the time-averaged Poynting vector over a surface $A_\mathrm{f}$ enclosing both the dipole and the sphere (see Figure~2(a)). Generalizing the analysis in \cite{elli2023strong} for time-varying media, $P_\mathrm{far}$ can be written as 

\begin{eqnarray}
P_\mathrm{far}=\frac{1}{2Z_0}\sum_{jl\mu s}{\frac{1}{k^2_j}|A^\mathrm{sca}_{jl\mu s}+d^\mathrm{(3)}_{jl\mu s}|^2}\label{eq:P_far}\,.
    \end{eqnarray}
    \noindent
Here, $Z_0$ is the vacuum impedance, $A^\mathrm{sca}_{jl\mu s}$ are the coefficients defined according to Equation~(6b) for the field scattered off the time-varying sphere. Note that to calculate $A^\mathrm{sca}_{jl\mu s}$, Equation~(1) is used with the incident field coefficients of the dipole given by Equation~(6a). Moreover, $d^\mathrm{(3)}_{jl\mu s}$ correspond to the coefficients of the dipole field defined in a basis of radiating VSHs (see Equation~\eqref{eq:dip_radiating}).

Further, to calculate the net power $P_\mathrm{LDOS}$ crossing a surface $A_\mathrm{s}$ enclosing only the dipole, we need to integrate the time-averaged Poynting vector of the total field over $A_\mathrm{s}$ (see Figure~2(a)). However, here, we outline an alternative approach to calculate $P_\mathrm{LDOS}$ as it is numerically easier to consider in our model. First, we write an expression for the instantaneous power $P_\mathrm{LDOS}(t)$ as \cite{Krasnok2015antenna,Yu2023manipulating,park2024spontaneous}
\begin{eqnarray}
P_\mathrm{LDOS}(t)=-\frac{1}{2}\textrm{Re}\left\{\int{dV_\mathrm{d}\,\left[\mathbf{J}_\mathrm{d}^*(\mathbf{r},t)\cdot\mathbf{E}^\mathrm{tot}(\mathbf{r},t)+\mathbf{J}_\mathrm{d}(\mathbf{r},t)\cdot\mathbf{E}^\mathrm{tot}(\mathbf{r},t)\right]}\right\}\label{eq:P_inst}\,.
    \end{eqnarray}  
    \noindent
Here, $V_\mathrm{d}$ is the volume enclosed by the surface $A_\mathrm{d}$ (see Figure~2(a)). Further, $\mathbf{E}^\mathrm{tot}(\mathbf{r},t)$ is the total field given by the sum of the field scattered from the time-varying sphere $\mathbf{E}^\mathrm{sca}(\mathbf{r},t)$ and the field of the dipole itself in the absence of the sphere $\mathbf{E}^\mathrm{dip}(\mathbf{r},t)$. In general, $\mathbf{E}^\mathrm{tot}(\mathbf{r},t)$ is polychromatic, i.e.,  $\mathbf{E}^\mathrm{tot}(\mathbf{r},t)=\sum_{j\in\mathbb{Z}}\mathbf{E}^\mathrm{tot}_j(\mathbf{r})e^{-i(\omega_\mathrm{d}+j\omega_\mathrm{m})t}$. Next, substituting the expressions for $\mathbf{J}_\mathrm{d}(\mathbf{r},t)$ and $\mathbf{E}^\mathrm{tot}(\mathbf{r},t)$ in Equation~\eqref{eq:P_inst} and taking the time-average, we can write $P_\mathrm{LDOS}$ as

\begin{eqnarray}
P_\mathrm{LDOS}=\langle P_\mathrm{LDOS}(t) \rangle=\frac{\omega_\mathrm{d}}{2}\textrm{Im}\left\{\mathbf{p}_0\cdot\mathbf{E}^\mathrm{tot}_0(\mathbf{r}_\mathrm{d})\right\}\,\label{eq:P_avg}.
    \end{eqnarray}
    \noindent
Note that the Equation~\eqref{eq:P_avg} is derived under the assumption that $\omega_\mathrm{d}\neq \frac{n \omega_\mathrm{m}}{2}$ with $n\in \mathbb{Z}$.

Finally, to calculate the net power $P_\mathrm{gain}-P_\mathrm{loss}$ crossing a surface $A_\mathrm{s}$ enclosing only the time-varying sphere, one needs to integrate the time-averaged Poynting vector over $A_\mathrm{s}$ (see Figure~2(a)). Generalizing the analysis in \cite{elli2023strong} for time-varying media, $P_\mathrm{gain}-P_\mathrm{loss}$ can be written as 

\begin{eqnarray}
P_\mathrm{gain}-P_\mathrm{loss}=\frac{1}{2Z_0}\mathlarger{\sum}_{jl\mu s}{\frac{1}{k^2_j}\left(|A^\mathrm{sca}_{jl\mu s}|^2+\textrm{Re}\left\{A^\mathrm{sca}_{jl\mu s}\cdot d^\mathrm{(1)*}_{jl\mu s}\right\}\right)}\label{P_net}\,.
    \end{eqnarray}
    \noindent
Here, $d^\mathrm{(1)}_{jl\mu s}$ correspond to the coefficients of the dipole field defined in a basis of regular VSHs (see Equation~\eqref{eq:dip_regular}).

\section{Explanation for time-averaging the observable quantities}
\begin{figure*}
\centerline{\includegraphics[width= 1\columnwidth]{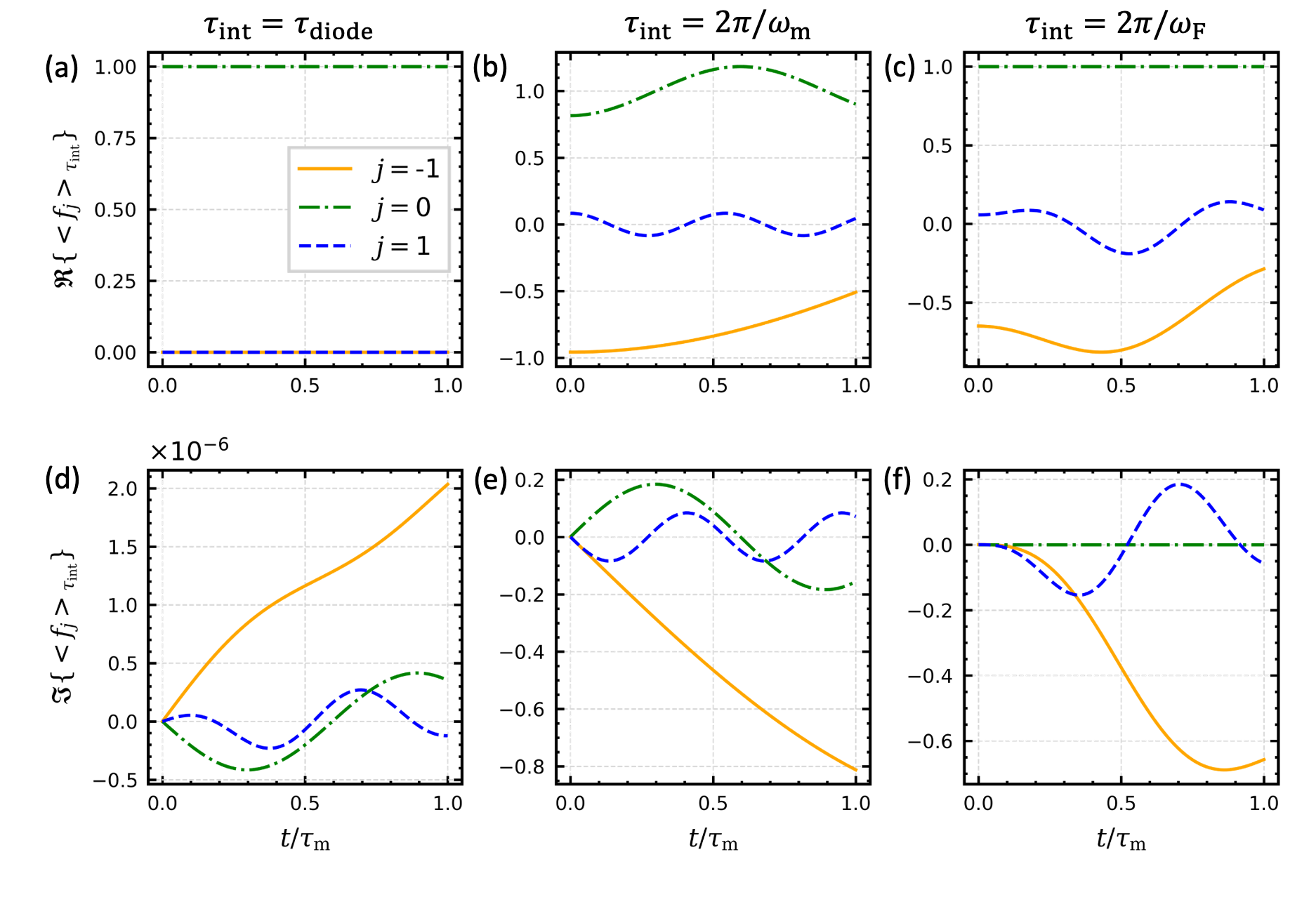}}
\caption{Real and imaginary part of $\langle f_j(t) \rangle_{\tau_\mathrm{int}}$ plotted as a function of time $t$. $\tau_\mathrm{int}=\tau_\mathrm{diode}$ for (a) and (d), $\tau_\mathrm{int}=2\pi/\omega_\mathrm{m}$ for (b) and (e) , and $\tau_\mathrm{int}=2\pi/\omega_\mathrm{F}$ for (c) and (f).}
\label{fig:avg}
\end{figure*}

Throughout the paper, we use time-averaged physical observables of interest. Here, we explain why such time-averaging is allowed even when the underlying scattering structure has time-varying material parameters. 

As an example, we consider a sphere-dipole system, as shown in Figure~3(d) of the main text. The instantaneous power $P_\mathrm{LDOS}(t)$ crossing a surface enclosing only the dipole is given by Equation~\eqref{eq:P_inst}. Further, the time-averaged power $<P_\mathrm{LDOS}(t)>_{\tau_\mathrm{int}}$ using the integration time $\tau_\mathrm{int}$ can be calculated as:
\begin{eqnarray}
<P_\mathrm{LDOS}(t)>_{\tau_\mathrm{int}}=\frac{1}{\tau_\mathrm{int}}\int_{t-\tau_\mathrm{int}/2}^{t+\tau_\mathrm{int}/2}P_\mathrm{LDOS}(t')\,dt'\,\label{eq:P_int}.
    \end{eqnarray}
    \noindent
Substituting Equation~\eqref{eq:P_inst} in Equation~\eqref{eq:P_int}, we get

\begin{subequations}\label{eq:P_int2}
\begin{align}
    \langle P_\mathrm{LDOS}(t) \rangle_{\tau_\mathrm{int}} &= \frac{\omega_\mathrm{d}}{2} \textrm{Im}\left\{ \mathbf{p}_0 \cdot \sum_{j \in \mathbb{Z}} \mathbf{E}_j^\mathrm{tot}(\mathbf{r}_\mathrm{d}) \langle f_j(t) \rangle_{\tau_\mathrm{int}} \right\} \label{eq:Pint3},\textrm{ where} \\
    \langle f_j(t) \rangle_{\tau_\mathrm{int}} &= e^{-ij\omega_\mathrm{m}t}\frac{\mathrm{sin}\left(\frac{j\omega_\mathrm{m}\tau_\mathrm{int}}{2}\right)}{\frac{j\omega_\mathrm{m}\tau_\mathrm{int}}{2}}-e^{-i(2\omega_\mathrm{F}+(j+2j_0)\omega_\mathrm{m})t}\frac{\mathrm{sin}\left(\frac{(2\omega_\mathrm{F}+(j+2j_0)\omega_\mathrm{m})\tau_\mathrm{int}}{2}\right)}{\frac{(2\omega_\mathrm{F}+(j+2j_0)\omega_\mathrm{m})\tau_\mathrm{int}}{2}} \label{eq:fj}\,.
\end{align}
\end{subequations}
Here, we have defined the Floquet frequency as $\omega_\mathrm{F}=\omega_\mathrm{d}-j_0\omega_\mathrm{m}$. Here, $j_0$ is an appropriately chosen positive integer such that $\omega_\mathrm{F}\in(0,\omega_\mathrm{m})$. For the results presented in this paper, we assumed $\tau_\mathrm{int}=\tau_\mathrm{diode}$. Here, $\tau_\mathrm{diode}$ is the integration time of a photodiode. Typically, $\tau_\mathrm{diode}$ is in the order of a few nanoseconds\cite{goushcha2017response}. Since the magnitudes of $\omega_\mathrm{F}$ and $\omega_\mathrm{m}$ are in the order of a few thousand terahertz (optical regime), we assume $\tau_\mathrm{diode}\gg2\pi/\omega_\mathrm{F},2\pi/\omega_\mathrm{m}$. Further, we also choose our system parameters such that $\omega_\mathrm{F}\neq\frac{n\omega_\mathrm{m}}{2}$ with $n\in\mathbb{Z}$. Such a choice is made as $\omega_\mathrm{F}=\frac{n\omega_\mathrm{m}}{2}$ corresponds to the case where the incident radiation gets coupled to the momentum bandgap modes of the time-varying nanostructure \cite{gaxiola2023growing}. Therefore, this leads to an exponential growth of the fields as a function of time. In such a scenario, the time-averaged analysis ceases to remain valid. Finally, we also assume that $|\omega_\mathrm{F}-n{\omega_\mathrm{m}}|\gg\frac{2\pi}{\tau_\mathrm{diode}}$ with $n\in\left\{0,\frac{1}{2},1\right\}$. Such a constraint is imposed as otherwise the second term in Equation~\eqref{eq:fj} leads to a time-dependent contribution to $\langle f_j(t) \rangle_{\tau_\mathrm{int}}$.

Under the assumptions mentioned above, we simplify Equation~\eqref{eq:fj} to
\begin{eqnarray}
\langle f_j(t) \rangle_{\tau_\mathrm{int}}\approx\begin{cases}1,\hspace{10pt}\mathrm{for}\hspace{6pt}j=0\,,\\
0,\hspace{10pt}\mathrm{for}\hspace{6pt}j\neq0\,.\label{eq:fjsimple}
\end{cases}
\end{eqnarray}
Finally, using Equation~\eqref{eq:fjsimple} and Equation~\eqref{eq:Pint3}, we write 
\begin{eqnarray}
\langle P_\mathrm{LDOS}(t) \rangle_{\tau_\mathrm{int}}= \frac{\omega_\mathrm{d}}{2} \textrm{Im}\left\{ \mathbf{p}_0 \cdot \mathbf{E}_0^\mathrm{tot}(\mathbf{r}_\mathrm{d})\right\}\,.\label{eq:Pldos_avg}
\end{eqnarray}
Note that the time-averaged $P_\mathrm{LDOS}(t)$ in Equation~\ref{eq:Pldos_avg} is identical to that in Equation~\ref{eq:P_avg}. Further, $\langle P_\mathrm{LDOS}(t) \rangle_{\tau_\mathrm{int}}$ is independent of time $t$. 

So far, we have assumed $\tau_\mathrm{int}=\tau_\mathrm{diode}$. Such a choice leads to time-independent physical observables upon averaging. However, in the considered system, there exist two more natural time scales, i.e., $\tau_\mathrm{int}=2\pi/\omega_\mathrm{m}$ and $\tau_\mathrm{int}=2\pi/\omega_\mathrm{F}$. Using $\tau_\mathrm{int}=2\pi/\omega_\mathrm{m}$ or $\tau_\mathrm{int}=2\pi/\omega_\mathrm{F}$ in Equation~\eqref{eq:P_int2}, does not necessarily lead to time-independent averaged quantities. In what follows, we show how choosing $\tau_\mathrm{int}$ differently affects $\langle P_\mathrm{LDOS}(t) \rangle_{\tau_\mathrm{int}}$. 

We assume the system parameters to be identical to those used for Figure~3(d) of the main text. In Figure~\ref{fig:avg}, we plot $\langle f_j(t) \rangle_{\tau_\mathrm{int}}$ for the cases where $\tau_\mathrm{int}\in\{\tau_\mathrm{diode},2\pi/\omega_\mathrm{m},2\pi/\omega_\mathrm{F}\}$. Further, we only show $\langle f_j(t) \rangle_{\tau_\mathrm{int}}$ for $j\in\{-1,0,1\}$. From Figure~\ref{fig:avg}(a) and (d), we observe that when $\tau_\mathrm{int}=\tau_\mathrm{diode}$, $\langle f_j(t) \rangle_{\tau_\mathrm{int}}$ satisfies Equation~\ref{eq:fjsimple}, leading to time-independent $\langle P_\mathrm{LDOS}(t) \rangle_{\tau_\mathrm{int}}$ (see Equation~\ref{eq:Pint3}). Whereas, from Figure~\ref{fig:avg}(b) and (e), we observe that $\langle f_j(t) \rangle_{\tau_\mathrm{int}}$ is dependent on $t$ when $\tau_\mathrm{int}=2\pi/\omega_\mathrm{m}$. Similarly, from  Figure~\ref{fig:avg}(c) and (f), we again observe that $\langle f_j(t) \rangle_{\tau_\mathrm{int}}$ depends on $t$ when $\tau_\mathrm{int}=2\pi/\omega_\mathrm{F}$. The time-dependent $\langle f_j(t) \rangle_{\tau_\mathrm{int}}$ leads to time-dependent $\langle P_\mathrm{LDOS}(t) \rangle_{\tau_\mathrm{int}}$ (see Equation~\eqref{eq:Pldos_avg}).

In summary, we conclude that $\langle P_\mathrm{LDOS}(t) \rangle_{\tau_\mathrm{int}}$ is independent of $t$ when $\tau_\mathrm{int}\gg2\pi/\omega_\mathrm{F},2\pi/\omega_\mathrm{m}$ in addition to the aforementioned assumptions. Furthermore, such a conclusion holds for all other time-averaged observables presented in this work.

\section{Spectral components of the total far field for the sphere-dipole system}
\begin{figure*}
\centerline{\includegraphics[width= 0.6\columnwidth]{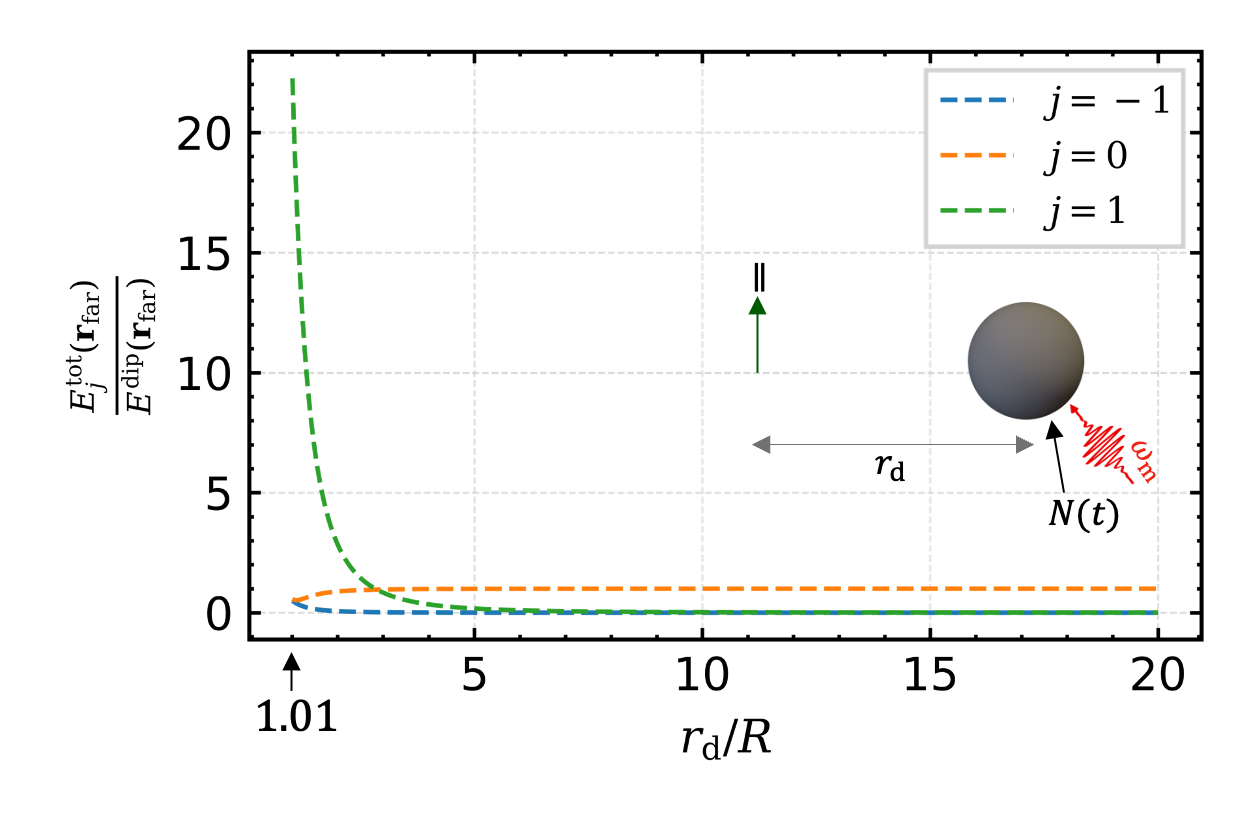}}
\caption{The normalized spectral components $\frac{E^\mathrm{tot}_j(\mathbf{r}_\mathrm{far})}{E^\mathrm{dip}(\mathbf{r}_\mathrm{far})}$ of the total far field as a function of the distance $r_\mathrm{d}$ of the dipole from the center of the optimized time-varying sphere.}
\label{fig:farfield}
\end{figure*}
\begin{figure*}
\centerline{\includegraphics[width= 0.8\columnwidth]{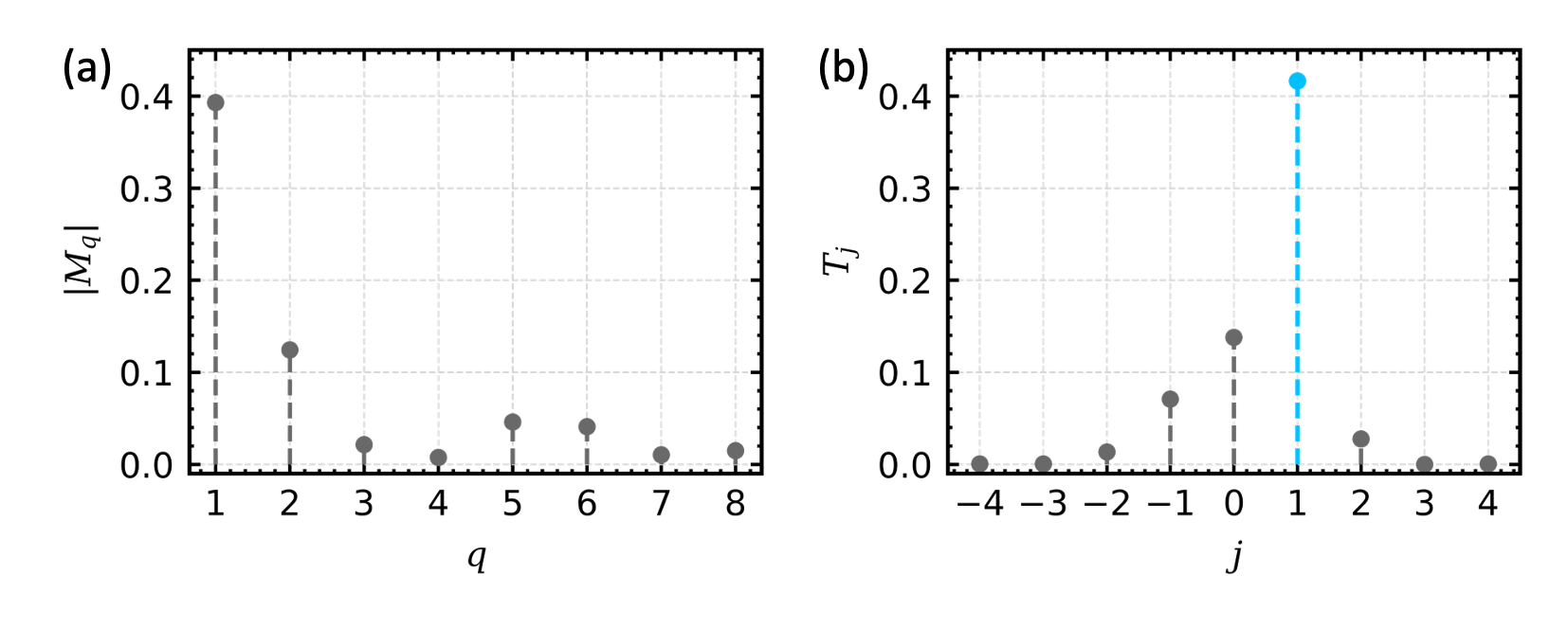}}
\caption{(a) The amplitudes of the Fourier components of the optimized electron density $N(t)$ of the metasurface shown in Figure~4(c). (b) The transmissivity contributions of the metasurface as a function of the harmonic number $j$ (identical to Figure~4(d)).}
\label{fig:fouriermetasurface}
\end{figure*}
In Figure~\ref{fig:farfield}, we show the normalized spectral components of the total far field $\mathbf{E}^\mathrm{tot}(\mathbf{r}_\mathrm{far})$ for the sphere-dipole system considered in Section~3 of the main text. Note that, we plot the spectral components for the tangential orientation of the dipole kept near the time-varying sphere with the optimized $N(t)$ shown in Figure~3(c) of the main text. Further, for simplicity, we only show the spectral components for $j\in\{-1,0,1\}$. From Figure~\ref{fig:farfield}, we observe that ${E}^\mathrm{tot}(\mathbf{r}_\mathrm{far})$ is maximized at $j=1$ for an appropriately chosen $N(t)$ when the dipole is kept at $r_\mathrm{d}=1.01R$. Such a maximization of ${E}^\mathrm{tot}(\mathbf{r}_\mathrm{far})$ at $j=1$ leads to the maximization of the far field power $P_\mathrm{far}$ of the sphere-dipole system, which in turn maximizes the decay rate enhancement $\gamma_\mathrm{rad}/\gamma_0$ of the dipole (see Equation~(5) and \eqref{eq:P_far}).

\section{Frequency conversion by spatiotemporal metasurfaces due to amplitude and phase modulation}\label{sec:phase-modulation}
For a periodically modulated spatiotemporal metasurface, the time-domain transmitted field $\mathbf{\Tilde{E}^\mathrm{tra}}(t)$ in response to an incident field $\mathbf{\Tilde{E}^\mathrm{inc}}(t)$ can be written as 
\begin{equation}
    \mathbf{\Tilde{E}^\mathrm{tra}}(t)=\Tilde{T}(t)\mathbf{\Tilde{E}^\mathrm{inc}}(t)\,,
\end{equation}
where $\Tilde{T}(t)$ is the time-domain complex transmission coefficient of the metasurface. Due to the periodic temporal modulation, $\Tilde{T}(t)=\sum_j\Tilde{T}_j(t)$, with $j\in\mathbb{Z}$. Here, $\Tilde{T}_j(t)=|\Tilde{T}_j(t)|e^{-ij\omega_\mathrm{m}t}$. Further, assuming a unit amplitude, $x$-polarized, and monochromatic incident plane wave $\mathbf{\Tilde{E}^\mathrm{inc}}(t)=e^{-i\omega_0t}\hat{\mathbf{x}}$ to the metasurface, the transmitted field can be written as
\begin{equation}
    \mathbf{\Tilde{E}^\mathrm{tra}}(t)=\sum_j|\Tilde{T}_j(t)|e^{-i(\omega_\mathrm{0}+j\omega_\mathrm{m})t}\,.
\end{equation}
\noindent However, to attain a perfect frequency upconversion to $\omega_\mathrm{0}+\omega_\mathrm{m}$, the transmitted field should be 
\begin{equation}
    \mathbf{\Tilde{E}^\mathrm{tra}}(t)=e^{-i(\omega_\mathrm{0}+\omega_\mathrm{m})t}\,.
\end{equation}
Such an upconversion is achieved if, $\Tilde{T}(t)=e^{-i\omega_\mathrm{m}t}$. In other words, the perfect upconversion to $\omega_\mathrm{0}+\omega_\mathrm{m}$ requires a pure phase modulation of the complex transmission coefficient $\Tilde{T}(t)$ of the nanostructures. Furthermore, this phase modulation should be such that the phase of $\Tilde{T}(t)$ goes from $0$ to $2\pi$ in a linear fashion as time $t$ goes from $0$ to $\tau_\mathrm{m}$ \cite{wu2020serrodyne}.

\section{Preferential spectral coupling for the spatiotemporal metasurface}
As shown in Figure~4(c) and (d) of the main text, an appropriately optimized electron density $N(t)$ leads to a preferential spectral coupling to $\omega_1$ at transmission. The amplitudes of the Fourier components of the optimized electron density are shown in Figure~\ref{fig:fouriermetasurface}(a). Further, the corresponding transmissivity contributions $T_j$  as a function of the harmonic number $j$ are shown in Figure~\ref{fig:fouriermetasurface}(b). Note that Figure~\ref{fig:fouriermetasurface}(b) is identical to Figure~4(d) of the main text. From Figure~\ref{fig:fouriermetasurface}(a), we observe that $M_1$ has the maximum amplitude among all $M_q$. This maximal $|M_1|$ along with the Mie resonance of the underlying static metasurface explains high $T_1$ in Figure~\ref{fig:fouriermetasurface}(b). Further, the maximal $|M_1|$ also causes a strong coupling to $T_{-1}$. However, since there is no Mie resonance at $\omega_{-1}$ of the underlying static metasurface, $T_{-1}$ remains significantly lower than $T_{1}$. Finally, from Figure~\ref{fig:fouriermetasurface}(b), we note that $T_{0}$ is also non-negligible. Such a non-negligible $T_{0}$ can be explained based on the fact that the spatiotemporal metasurface is unable to provide a pure linear phase modulation of $2\pi$ of the time-domain transmission coefficient $\Tilde{T}(t)$ (see Section~\ref{sec:phase-modulation}).

\section{Fabry-Perot cavity based filter}
The narrow-band filter shown in Figure~4(a) of the main text can be implemented using a Fabry-Perot cavity. We consider a cavity formed by two plane-parallel mirrors. The reflectivity of each mirror is assumed to be $R=0.95$. Note that such mirrors can be realized by using, for instance, Bragg stacks \cite{Joannopoulos2008photonic}. Further, the distance between the mirrors is $d=213$~nm. Moreover, we assume the space between the mirrors to be filled with vacuum. Using the parameters mentioned above, we obtain a narrow-band filter whose transmissivity profile is shown in Figure~4(b).

\section{Test for nonreciprocity}
\begin{figure*}
\centerline{\includegraphics[width= 0.8\columnwidth]{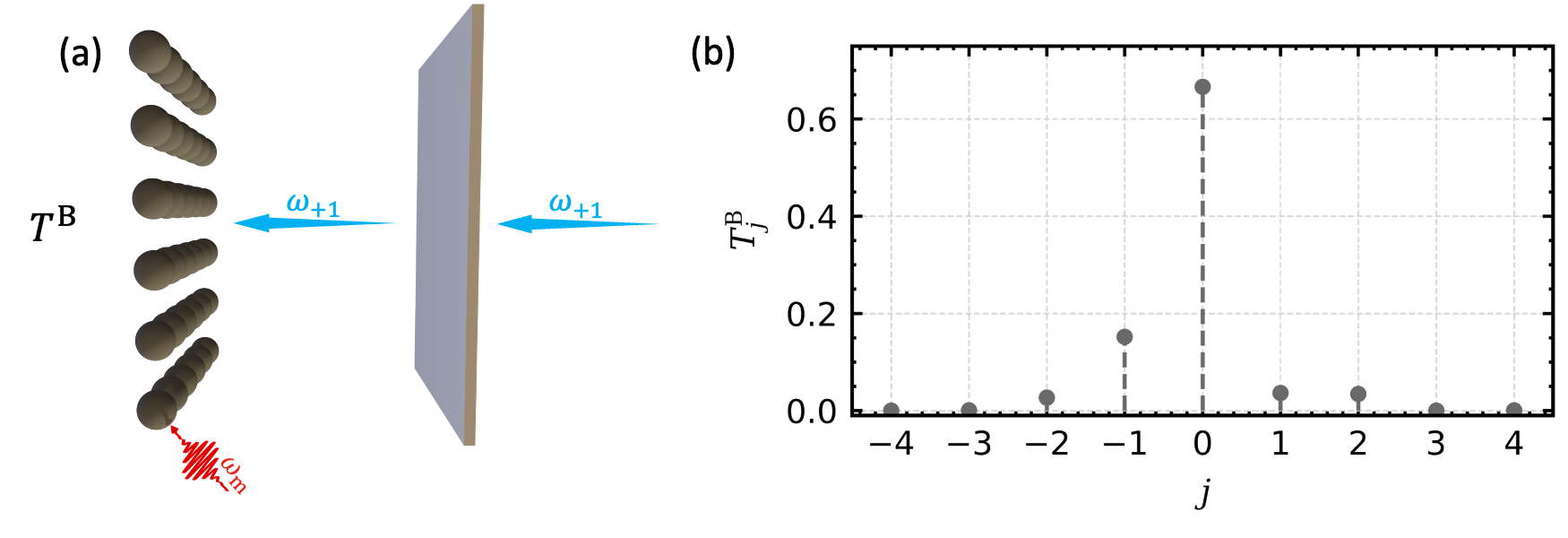}}
\caption{Testing nonreciprocity. (a) The cascaded system formed by a spatiotemporal metasurface and a filter (identical to that in Figure~4(a) of the main text). Here, we assume a monochromatic $x$-polarized plane wave at frequency $\omega_1$ normally incident to the filter. (b) The backward transmissivity contributions of the cascaded system for the given incident field as a function of the harmonic number $j$.}
\label{fig:nr}
\end{figure*}

In the main text, we demonstrated that the system shown in Figure~4(a) supports asymmetric transmission (AT). In this section, we test if the same system also exhibits nonreciprocity. To perform such a test, we assume a monochromatic x-polarized plane wave at frequency $\omega_1$ normally incident to the filter (see Figure~\ref{fig:nr}(a)). Further, we show the backward transmissivity contributions $T^\mathrm{B}_j$ for such an incident field after its propagation through the cascaded system (see Figure~\ref{fig:nr}(b)). Note that, according to \cite[Equation~(118)]{asadchy2020tutorial}, the considered time-varying system is reciprocal if it satisfies the generalized Lorentz reciprocity condition given by 
\begin{eqnarray}
\frac{E^\mathrm{F}_1}{E^\mathrm{B}_0}=\frac{\omega_1}{\omega_0}\,.\label{eq:nr1}
\end{eqnarray}
Here, $E^\mathrm{F}_1$ corresponds to the complex amplitude of the forward transmitted field through the cascaded system at the frequency $\omega_1$ for an incident plane wave of unit amplitude at frequency $\omega_0$ (see Figure~4(a)). Similarly, $E^\mathrm{B}_0$ corresponds to the complex amplitude of the backward transmitted field through the cascaded system at the frequency $\omega_0$ for an incident plane wave of unit amplitude at frequency $\omega_1$ (see Figure~\ref{fig:nr}(a)). Next, using $T^\mathrm{F}_1=|E^\mathrm{F}_1|^2$ and $T^\mathrm{B}_0=|E^\mathrm{B}_0|^2$, we arrive at a \textit{necessary} condition that must be satisfied by our system to be reciprocal, given by
\begin{eqnarray}
\frac{T^\mathrm{F}_1}{T^\mathrm{B}_0}=\left(\frac{\omega_1}{\omega_0}\right)^2\,.\label{eq:nr2}
\end{eqnarray}
Here, $T^\mathrm{F}_1$ corresponds to the forward transmissivity contribution of the cascaded system for an incident field at frequency $\omega_0$ (see Figure~4(a)). Next, we substitute $T^\mathrm{F}_1=0.41531$, $T^\mathrm{B}_0=0.6662$, $\omega_0=2\pi \times 542$~THz, and $\omega_1=2\pi \times 704$~THz in Equation~\eqref{eq:nr2}. We find $\frac{T^\mathrm{F}_1}{T^\mathrm{B}_0}=0.62$, and $\left(\frac{\omega_1}{\omega_0}\right)^2=1.69$. Since our cascaded system does not satisfy Equation~\eqref{eq:nr2}, we conclude that it is nonreciprocal. 

The origin of the nonreciprocity of our system can be explained as follows: time-varying systems that are reciprocal must possess a generalized time-reversal symmetry about any arbitrary moment in time $t_0$ (see \cite[Equation~(11)]{williamson2020integrated}). Our optimized electron density $N(t)$ shown in Figure~4(c) breaks the time-reversal symmetry. Therefore, the considered cascaded system exhibits nonreciprocity. Note that another trivial $N(t)$ profile known to break the time-reversal symmetry and hence reciprocity corresponds to a sawtooth-type modulation \cite{williamson2020integrated,liu2018huygens}.

\section{Details of the convergence parameters used during simulations}
For performing the simulations throughout the article, we used various convergence parameters. These parameters are the maximum multipolar order $l_\mathrm{max}$, the total number of scattered frequencies $J$, the total number of Fourier coefficients $Q$ of $N(t)$, and the threshold $\alpha$ for the penalty function. Here, we list the actual numerical values of these parameters for the reproducibility of our results.

\noindent Figure~2(b): $l_\mathrm{max}=8,\,\textrm{and }J=1$.

\noindent Figure~3(a): $l_\mathrm{max}=5,\,\textrm{and }J=1$.

\noindent Figure~3(b): $l_\mathrm{max}=1,\,\textrm{and }J=1$.

\noindent Figure~3(c): $Q=6$.

\noindent Figure~3(d): For the backward simulation, we used $l_\mathrm{max}=3,\,J=11,\,Q=6,\,\textrm{and }\alpha=0.1$. On the other hand, for the forward simulation, we used $l_\mathrm{max}=8,\,J=61$.

\noindent Figure~4(b): $l_\mathrm{max}=5,\,\textrm{and }J=1$.

\noindent Figure~4(c): $Q=8$.

\noindent Figure~4(d): For the backward simulation, we used $l_\mathrm{max}=3,\,J=7,\,Q=8,\,\textrm{and }\alpha=0.055$. On the other hand, for the forward simulation, we used $l_\mathrm{max}=8,\,\textrm{and }J=61$.

\end{document}